# Measuring the knot of non-Hermitian degeneracies and non-commuting braids

Y. S. S. Patil[1*], J. Höller[1,2], P. A. Henry[3], C. Guria[1], Y. Zhang[1], L. Jiang[1], N. Kralj[1,4], N. Read[1,3,5], J. G. E. Harris[1,3,5*]

[1] *Department of Physics, Yale University, New Haven, Connecticut 06520, USA*
[2] *Howard Hughes Medical Institute, Janelia Research Campus, Ashburn, Virginia 20147, USA*
[3] *Department of Applied Physics, Yale University, New Haven, Connecticut 06520, USA*
[4] *Niels Bohr Institute, University of Copenhagen, Blegdamsvej 17, 2100 Copenhagen, Denmark*
[5] *Yale Quantum Institute, Yale University, New Haven, Connecticut 06520, USA*
* *corresponding authors: yogesh.patil@yale.edu, jack.harris@yale.edu*

**Any system of coupled oscillators may be characterized by its spectrum of resonance frequencies (or eigenfrequencies), which can be tuned by varying the system's parameters. The relationship between control parameters and the eigenfrequency spectrum is central to a range of applications.[1,2,3] However, fundamental aspects of this relationship remain poorly understood. For example, if the controls are varied along a path that returns to its starting point (i.e., around a "loop"), the system's spectrum must return to itself. In systems that are Hermitian (i.e., lossless and reciprocal) this process is trivial, and each resonance frequency returns to its original value. However, in non-Hermitian systems, where the eigenfrequencies are complex, the spectrum may return to itself in a topologically non-trivial manner, a phenomenon known as spectral flow. The spectral flow is determined by how the control loop encircles degeneracies, and this relationship is well understood for $N = 2$ (where $N$ is the number of oscillators in the system).[4,5] Here we extend this description to arbitrary $N$. We show that control loops generically produce braids of eigenfrequencies, and for $N > 2$ these braids form a non-Abelian group which reflects the non-trivial geometry of the space of degeneracies. We demonstrate these features experimentally for $N = 3$ using a cavity optomechanical system.**

A very wide range of physical systems are described by first-order differential equations of motion that are linear in the system's coordinates. This includes classical systems near to mechanical equilibrium (for example, coupled oscillators and linear wave systems), closed quantum systems, and open quantum systems that can be brought to Lindblad form. In these descriptions, the system's state is an $N$-dimensional complex vector whose time evolution is generated by an $N \times N$ complex matrix $\mathbf{H}$ (which we take to be traceless without loss of generality). The qualitative behavior of such a system depends on the form of $\mathbf{H}$, which reflects the relevant symmetries and conservation laws. For example, in the quantum description of closed systems, $\mathbf{H}$ is Hermitian. On the other hand, Newtonian mechanics and Maxwellian electromagnetism both allow for linear elements having nonreciprocity, gain, and loss, and so the classical equations of motion for $N$ coupled oscillators (whose positions and momenta are encoded as $N$ complex numbers) may have $\mathbf{H}$ of any form.

Recent years have seen considerable interest in features that distinguish non-Hermitian systems from their Hermitian counterparts. These include non-orthogonal eigenvectors; complex eigenvalues; and a type of degeneracy, known as an exceptional point (EP), at which $\mathbf{H}$ is non-diagonalizable. In addition, non-Hermitian systems respond to perturbations of $\mathbf{H}$ in a qualitatively different manner than Hermitian systems do.[4,6,7] These differences offer practical routes to new forms of control, sensing, and robustness, and have been explored in optics, microwaves, electronics, acoustics, optomechanics, and qubits. [1,2,3,5,8,9,10,11,12,13,14,15]

Despite rapid progress, some fundamental aspects of non-Hermitian systems remain poorly understood. For example, when a system's parameters are varied around a closed loop (with this "control loop" chosen so that the spectrum is non-degenerate throughout), the eigenvalues may move around one another in the complex plane. The way in which they do so, viewed topologically, is what we will describe below as "spectral flow". It is determined by the manner in which the control loop encloses degeneracies; however, the specific relationship between the loop, the degeneracies, and the resulting spectral flow is well known only for $N = 2$. For $N > 2$, studies of spectral flow have focused on special cases in which $\mathbf{H}$ is constrained, or on numerical simulations of specific systems, rather than on a general description of the spectral flow.[16,17,18,19,20,21,22,23,24]

## Control loops and spectral flow

For any $N$, the spectral flow can be described by regarding the spectrum of $\mathbf{H}$ as an unordered set $\boldsymbol{\lambda}$ of $N$ points in the complex plane. We take the parameters controlling $\boldsymbol{\lambda}$ to be the $N-1$ complex coefficients in $p_{\mathbf{H}}$, the characteristic polynomial of $\mathbf{H}$. These coefficients define the "control space" $\mathcal{L}_N \cong \mathbb{C}^{N-1}$. They smoothly parameterize the space of spectra, and have simple expressions in terms of the elements of $\mathbf{H}$. $\mathcal{L}_N$ can be partitioned into two subspaces $\mathcal{V}_N$ and $\mathcal{G}_N$, corresponding respectively to whether or not the spectrum is degenerate. $\mathcal{V}_N$ consists of the points where $D$, the discriminant of $p_{\mathbf{H}}$, vanishes (see Methods).

Although $\mathcal{L}_N$ is topologically trivial, this need not be the case for $\mathcal{G}_N$ and $\mathcal{V}_N$. To describe these two subspaces, we note that varying the control parameters along a smooth curve $C \subset \mathcal{G}_N$ causes the $N$ points in $\boldsymbol{\lambda}$ to be smoothly transported in the complex plane.[4] Throughout, we take $C$ to be a closed curve (or "loop"); we also fix a basepoint in $\mathcal{G}_N$ and consider only loops starting and ending at that point. In this case, traversing a loop $C$ causes the initial spectrum to be smoothly transported back to itself. Such an evolution of $N$ distinct points in the complex plane is called a braid of $N$ strands (for example, Ref. [25]). We will say that two braids are isotopic to one another if one of them can be continuously deformed into the other, keeping its endpoints fixed and its strands non-intersecting during the deformation. We define the spectral flow produced by a control loop $C$ to be the isotopy equivalence class $b$ of the corresponding braid of eigenvalues. Braids with the common basepoint can be concatenated to produce another such braid, and with that operation the $b$s form a group $B_N$, the Artin braid group[26].

Two isotopic braids arise from two control loops $C_1$, $C_2$ that can be continuously deformed into each other within $\mathcal{G}_N$, and hence each $b$ corresponds to a homotopy class[27] $\ell$ of based loops $C \subset \mathcal{G}_N$. Concatenating $C$s gives a group operation on the $\ell$s, which thus form the fundamental group[27] $\pi_1$ of the space $\mathcal{G}_N$. It follows from this discussion that $\pi_1(\mathcal{G}_N) \cong B_N$.[28,29,30] Because $\mathcal{L}_N$ is topologically trivial, the nontrivial $\pi_1(\mathcal{G}_N)$ arises solely because $\mathcal{V}_N$ (consisting of the points at which the spectrum is degenerate) was removed from $\mathcal{L}_N$ to produce $\mathcal{G}_N$, leaving a hole that control loops can wind around in various (non-homotopic) ways that correspond to the elements of $\pi_1$.

To give a concrete picture of $\mathcal{G}_N$ and $\mathcal{V}_N$ (and the ways in which loops in the former may encircle the latter), we note that for $N = 2$, the reasoning above returns the familiar result $\mathcal{G}_2 \cong \mathbb{C}\backslash\{0\}$ (the complex plane without the origin). The fundamental group of this space, $\pi_1(\mathcal{G}_2)$, is isomorphic to $B_2 \cong \mathbb{Z}$ (the group of integers under addition), reflecting the fact that each loop in $\mathcal{G}_2$ belongs to the $\ell$ determined by its winding number, and concatenating loops results in a new loop whose winding number is the sum of the winding numbers of the concatenated loops.

For $N = 3$, we have $\mathcal{G}_3 \cong \mathbb{C}^2 - \mathcal{V}_3$ and $\pi_1(\mathbb{C}^2 - \mathcal{V}_3) \cong B_3$. From the equation $D = 0$ we show in Methods that $\mathcal{V}_3$ is a connected hypersurface that includes a singular point at the origin (0,0) corresponding to three-fold degeneracy; the rest of $\mathcal{V}_3$ consists of the two-fold degeneracies. The two-fold degeneracies form the space $\mathcal{K} \times \mathbb{R}_{>0}$, where $\mathcal{K}$ is the trefoil knot and $\mathbb{R}_{>0}$ plays the role of the radial distance from the three-fold degeneracy. Therefore, if we identify $\mathbb{C}^2$ with $\mathbb{R}^4$, intersecting $\mathcal{V}_3$ with a real hypersphere $\mathcal{S}^3$ centered at the origin gives $\mathcal{K}$. This structure (which is shown in Extended Data Fig. 1) agrees with the fact that $\pi_1(\mathcal{S}^3 - \mathcal{K}) \cong B_3$.

This description highlights two important features common to all non-Hermitian systems with $N > 2$, but absent in the well-studied case $N = 2$. The first is that the subspace $\mathcal{V}_N$ has a non-trivial geometry. The second is that this geometry makes loops in $\mathcal{G}_N$ non-commutative (as $B_N$ is non-Abelian for $N > 2$). This rich behavior reflects the fact that $\boldsymbol{\lambda}$ consists of the roots of $p_{\mathbf{H}}$, and non-Hermitian systems can realize any complex polynomial as $p_{\mathbf{H}}$. In the mathematical context of complex polynomial equations, the braid and knot structures described here are well-known features of the relation between a polynomial's coefficients and its roots.

Here, we provide an experimental demonstration of these two features. We use a three-mode mechanical system in which $\mathbf{H}$ is tuned by control parameters $\boldsymbol{\Psi}$ that span $\mathcal{L}_3$ and so provide access to a three-fold degeneracy and all the spectra in its neighborhood.[6] We measure spectra on a hypersurface surrounding the three-fold degeneracy, and find the trefoil knot $\mathcal{K}$ formed by the two-fold degeneracies. We show that varying $\boldsymbol{\Psi}$ around a loop produces an eigenvalue braid whose spectral flow is determined by how the loop encircles $\mathcal{K}$. We demonstrate braids that can be concatenated to produce any element of $B_3$, and show the non-commutation of these braids. These features are demonstrated using a cavity optomechanical system, though we emphasize that they are generic to oscillators realized in any physical domain.

**Experimental system**

The experiment is shown schematically in Fig. 1a. It uses three vibrational modes of a $Si_3N_4$ membrane. The dynamical matrix $\mathbf{H}$ governing these modes is controlled by placing the membrane in an optical cavity and using the dynamical back-action (DBA) effect of cavity optomechanics.[31] In the absence of DBA, the three modes have resonance frequencies $(\tilde{\omega}_1^{(0)}, \tilde{\omega}_2^{(0)}, \tilde{\omega}_3^{(0)}) =$ $2\pi \times (352{,}557{,}705)$ kHz and optomechanical coupling rates $\boldsymbol{g} = (g_1, g_2, g_3) = 2\pi \times (0.198, 0.304, 0.300)$ Hz. The cavity linewidth $\kappa = 2\pi \times 190$ kHz. Three tones produced from a single laser ("control" in Fig. 1a) drive the cavity with powers $P_{1,2,3}$. The tones' relative detunings are fixed (Fig. 1b), and their beatnotes define a rotating frame $\mathcal{R}$ in which the three modes' eigenvalues are almost degenerate for $P_i = 0$. Within $\mathcal{R}$, the control parameters $\boldsymbol{\Psi} = (\delta, P_1, P_2, P_3)$ (where $\delta$ is the tones' common detuning, Fig. 1b) can tune the system to a three-fold degeneracy. They also provide linearly independent control of the coefficients of $p_{\mathbf{H}}$ (see Methods and Supplementary Information), and hence span $\mathcal{L}_3$. $\mathbf{H}$ is otherwise unconstrained, so it accesses degeneracies of the most generic type (for a given order): i.e., at an $m$-order degeneracy, the Jordan normal form of $\mathbf{H}$ contains a Jordan block of dimension $m$ (we call such a point $\mathrm{EP}_m$).

The modes' eigenvalue spectrum $\boldsymbol{\lambda}$ is determined by measuring the membrane's mechanical susceptibility. This is accomplished using a second laser ("probe" in Fig. 1a) to exert an oscillatory force on the membrane (at frequency $\tilde{\omega}_{\mathrm{AM}}$), and to record a heterodyne signal $\tilde{V}$ proportional to the membrane's response. Fig. 1c shows a typical measurement of $\tilde{V}(\tilde{\omega}_{\mathrm{AM}})$, along with a fit of this data to standard optomechanics theory. This fit returns the complex eigenvalues $\lambda_i$, as well as the amplitudes $s_{i,j}$ (denoting the contribution of the $j^{\text{th}}$ mode to the peak near $\tilde{\omega}_i^{(0)}$). In the remainder of this paper $\boldsymbol{\lambda}$ is determined from data and fits as in Fig. 1c.

**Locating degeneracies**

The system's $\mathrm{EP}_3$ is identified by measuring $\boldsymbol{\lambda}(\boldsymbol{\Psi})$ and converting each $\boldsymbol{\lambda}$ to $d = |\lambda_1 - \lambda_2| + |\lambda_2 - \lambda_3| + |\lambda_3 - \lambda_1|$ (see Methods). As shown in Extended Data Figs. 2,3, measurements of $d(\boldsymbol{\Psi})$ give $\boldsymbol{\Psi}_{\mathrm{EP3}} = (2\pi \times 54(7)\ \mathrm{kHz}, 128(8)\ \mu\mathrm{W}, 428(3)\ \mu\mathrm{W}, 304(15)\ \mu\mathrm{W})$, in good agreement with the value calculated from the independently-measured device parameters (see Methods).

To study the spectrum on a hypersurface surrounding $\boldsymbol{\Psi}_{\mathrm{EP3}}$, we measured $\boldsymbol{\lambda}$ on the boundary of a 4D hyperrectangle $\mathcal{S}$ centered close to $\boldsymbol{\Psi}_{\mathrm{EP3}}$. Specifically, $\mathcal{S}$ bounds the region: $-10\ \mathrm{kHz} \leq \delta/2\pi \leq 106\ \mathrm{kHz}$, $22\ \mu\mathrm{W} \leq P_1 \leq 240\ \mu\mathrm{W}$, $289\ \mu\mathrm{W} \leq P_2 \leq 675\ \mu\mathrm{W}$, $78\ \mu\mathrm{W} \leq P_3 \leq 702\ \mu\mathrm{W}$. It consists of eight 3D "faces", each corresponding to fixing the value of one control parameter. $\boldsymbol{\Psi}$ was densely rastered over 61 distinct 2D "sheets" within $\mathcal{S}$ (Extended Data Fig. 4). Data from a typical sheet is shown in Fig. 2. For each value of $\boldsymbol{\Psi}$ (i.e., for each pixel in the sheet), $\tilde{V}(\tilde{\omega}_{\mathrm{AM}})$ was measured and fit as in Fig. 1c.

To locate the $\mathrm{EP}_2$ points in $\mathcal{S}$ we considered two quantities derived from these fits: $D = (\lambda_1 - \lambda_2)^2(\lambda_2 - \lambda_3)^2(\lambda_3 - \lambda_1)^2$ and $E = (\det[\mathbf{S}])^{-2}$ where $\mathbf{S}$ is the matrix formed by the $s_{i,j}$. Both $D$ and $E$ vanish at $\mathrm{EP}_2$, and both exhibit a phase winding of $2\pi$ around $\mathrm{EP}_2$. However, they provide complementary information: $D = 0$ reflects eigenvalue degeneracy, while $E = 0$ reflects eigenvector degeneracy (see Supplementary Information). Furthermore, $D$ and $E$ are derived from different aspects of the fits to $\tilde{V}(\tilde{\omega}_{\mathrm{AM}})$, and so reflect partially independent features of the data. The locations ($\boldsymbol{\Psi}_{\mathrm{EP2}}$) of the zeros and phase windings in $D$ and $E$ are identified algorithmically (Supplementary Information) and are shown in Fig. 2 as cyan circles.

**The knot of two-fold degeneracies**

Figures 3a,b show all of the $\boldsymbol{\Psi}_{\text{EP2}}$ identified in this way. For ease of visualization, they are depicted using two projections of $\mathcal{S}$, both of which generically preserve knot equivalence classes. Fig. 3a uses a stereographic projection, while Fig. 3b uses a projection isomorphic to the one in Fig. 3a, but which is more easily connected to the control parameters. In both projections, the experimentally identified $\text{EP}_2$'s are seen to trace out a curve that forms a trefoil knot $\mathcal{K}$. Each point in Figs. 3a,b is colored according to the value of $\theta$ measured at the corresponding $\boldsymbol{\Psi}_{\text{EP2}}$ ($\theta$ is derived from $\boldsymbol{\lambda}$ as defined in Sec. 7 of the Supplementary Information, and serves as a coordinate along $\mathcal{K}$).

Figures 3a,b also show the best fit of the measured $\boldsymbol{\Psi}_{\text{EP2}}$ to standard optomechanics theory (see Methods). This fit uses $\boldsymbol{g}$ and $\kappa$ as parameters and returns $\boldsymbol{g} = 2\pi \times (0.1979, 0.3442, 0.3092)$ Hz and $\kappa = 2\pi \times 173.84$ kHz (these values are also used to generate the plots of $D$ and $E$ (labelled "theory") in the right-hand column of Fig. 2). These values of $\boldsymbol{g}$ and $\kappa$ extracted by fitting the knot $\mathcal{K}$ in the three-mode spectrum agree well with the values given earlier (in the section 'Experimental system'), which are determined independently from measurements of the DBA (see Extended Data Fig. 9 and Supplementary Information).

**Non-commuting eigenvalue braids**

When $\boldsymbol{\Psi}$ is varied around a loop $C$ from a given $\ell$, $\boldsymbol{\lambda}(\boldsymbol{\Psi})$ is expected to form a braid whose equivalence class $b$ is determined by $\ell$. To demonstrate this, we selected pixels from the dataset described above (the 61 sheets) that trace out three loops with a common basepoint, as shown in Figs. 3c-e. The corresponding $\boldsymbol{\lambda}(\boldsymbol{\Psi})$ for each loop is shown in Figs. 3f-h. The loops belong to different $\ell$, and result in eigenvalue braids from $b = \mathbb{I}, \sigma_1, \sigma_2\sigma_1$ (Figs. 3f-h, respectively). Here, $\mathbb{I}$ is the identity, $\sigma_i$ ($\sigma_i^{-1}$) indicates that strand $i$ has crossed over (under) strand $i+1$, the strands are counted from the left (in the view used for the figures), and operations are written symbolically from right to left as the braid is read from bottom to top.[25] Since $\sigma_1$ and $\sigma_2\sigma_1$ together generate the group $B_3$, the loops in Figs. 3d,e can be concatenated to produce any braid of eigenvalues. The correspondence between a loop's $\ell$ and the $b$ it produces is a robust feature of the data; this is illustrated in Fig. 4 and Extended Data Fig. 7, which shows the braids produced by a number of other loops.

The non-Abelian character of the group formed by these braids is demonstrated in Fig. 4. Fig. 4a shows two loops (red, blue) belonging to different $\ell$. Fig. 4b shows $\boldsymbol{\lambda}(\boldsymbol{\Psi})$ as $\boldsymbol{\Psi}$ is stepped first around the red loop and then around the blue loop, while Fig. 4c shows $\boldsymbol{\lambda}(\boldsymbol{\Psi})$ as $\boldsymbol{\Psi}$ is stepped first around the blue loop and then around the red loop. The former gives $b = \sigma_2^{-1}\sigma_1^{-1}$, while the latter gives $b = \sigma_1^{-1}\sigma_2^{-1}$. The inequivalence of these braids, which can be seen directly from the fact that they result in different permutations of the eigenvalues, demonstrates that encircling a degeneracy is not characterized by a number (as is the case for $N = 2$), but by a braid equivalence class.

**Future directions**

Looking ahead, one may ask if the braids demonstrated here may play a role in the system's dynamics. For example, if one eigenmode of the system is initially excited, and then the system is slowly evolved around a control loop, it might be expected that the excitation would remain in the eigenmode that is smoothly connected with the original one, in analogy with adiabatic transport in Hermitian systems. If this were the case, a control loop would permute excitations among the normal modes, with the specific permutation determined by the loop's $\ell$. Such a control scheme – where the outcome is determined by a topological property of the input – would be of considerable interest. However, in non-Hermitian systems adiabatic control loops do not transport excitations in this manner.[32] On the other hand, real-time loops have been shown to produce similar transport in special cases,[33,34,13,14] and it remains an open question whether control schemes such as "shortcuts to adiabaticity"[35,36,37] or tailored nonlinearities[38,39,12,40] can stabilize such transport more generally. Exploration of these possibilities may open new means for achieving robust topological control in oscillator systems.

[30] Arnold, V. I. “On some topological invariants of algebraic functions” in *Vladimir I. Arnold, Collected Works Volume II*, A. B. Givental, B. A. Khesin, A. N. Varchenko, A. A. Vassiliev, O. Ya. Viro Eds. (Springer-Verlag, Berlin Heidelberg, 2014) pp. 199-220.

[31] Aspelmeyer, M., Kippenberg, T. J., Marquardt, F. Cavity optomechanics, *Reviews of Modern Physics* **86**, 1391 – 1452 (2014).

[32] Nenciu, G. & Rasche, G. On the adiabatic theorem for nonself-adjoint Hamiltonians, *Journal of Physics A* **25**, 5741 (1992).

[33] Uzdin, R., Mailybaev, A. & Moiseyev, N. On the observability and asymmetry of adiabatic state flips generated by exceptional points, *Journal of Physics A* **44**, 435302 (2011).

[34] Berry, M. V. & Uzdin, R. Slow non-Hermitian cycling: exact solutions and the Stokes phenomenon, *Journal of Physics A* **44**, 435303 (2011).

[35] Emmanouilidou, A., Zhao, X. G., Ao, P. & Niu, Q. Steering an Eigenstate to a Destination, *Physical Review Letters* **85**, 1626 (2000).

[36] Berry, M. V. Transitionless quantum driving, *Journal of Physics A* **42**, 365303 (2009).

[37] Ibáñez, S., Martínez-Garaot, S., Chen, X., Torrontegui, E. & Muga, J. G. Shortcuts to adiabaticity for non-Hermitian systems, *Physical Review A* **84**, 023415 (2011).

[38] Wu, B., Liu, J. & Niu, Q. Geometric Phase for Adiabatic Evolutions of General Quantum States, *Physical Review Letters* **94**, 140402 (2005).

[39] Graefe, E.-M. & Korsch, H. J. Crossing scenario for a nonlinear non-Hermitian two-level system, *Czechoslovak Journal of Physics* **56**, 1007–1020 (2006).

[40] Wang, H., Assawaworrarit, S. & Fan, S. Dynamics for encircling an exceptional point in a nonlinear non-Hermitian system, *Optics Letters* **44**, 638 – 641 (2019).

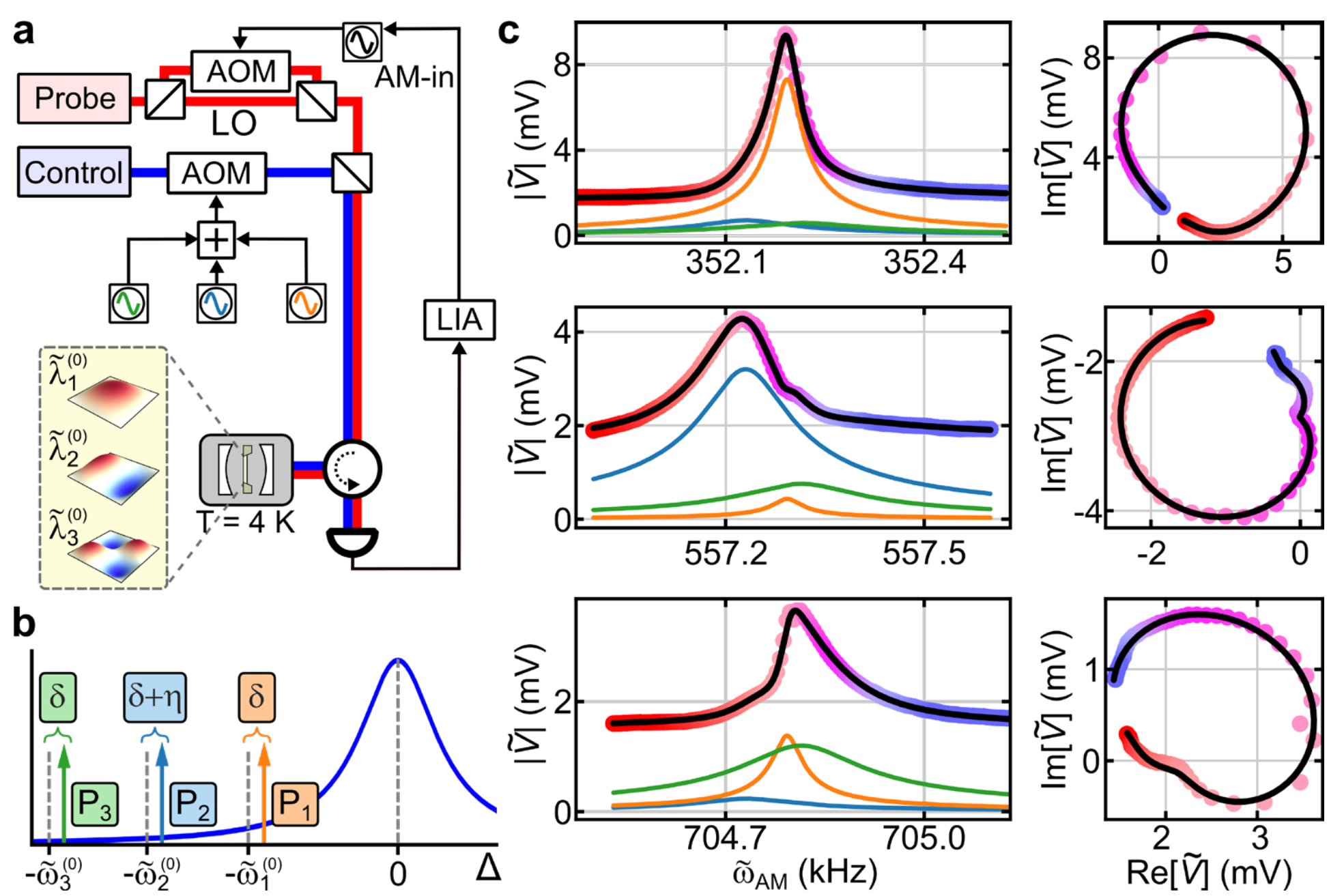

**Fig. 1 | Experimental schematic and susceptibility measurement. a,** Two lasers (red and blue paths) address two modes of an optical cavity (white) in a cryostat (gray). The cavity contains a $Si_3N_4$ membrane (yellow) whose three mechanical modes are shown in the dashed box. AOM: acousto-optic modulator; LO: local oscillator; LIA: lock-in amplifier; AM-in: amplitude modulation input. **b,** The optical spectrum, showing the three control beams (green, light blue, orange) and the cavity mode (blue). The non-degeneracy of the bare modes in $\mathcal{R}$ is set by $\eta = -2\pi \times 100$ Hz. **c,** The complex response $\tilde{V}$ measured at frequencies $\tilde{\omega}_{\mathrm{AM}}$ near $\tilde{\omega}_1^{(0)}$ (top), $\tilde{\omega}_2^{(0)}$ (center) and $\tilde{\omega}_3^{(0)}$ (bottom). Here $\boldsymbol{\Psi} = (2\pi \times 50\ \mathrm{kHz}, 125\ \mu W, 364\ \mu W, 426\ \mu W)$. The left column shows $|\tilde{V}(\tilde{\omega}_{\mathrm{AM}})|$ and the right column shows a parametric plot of $\tilde{V}(\tilde{\omega}_{\mathrm{AM}})$. The data points are colored by $\tilde{\omega}_{\mathrm{AM}}$ (the 1-$\sigma$ confidence intervals are smaller than the plotted points). A global fit (black lines) gives the system's eigenvalues. The magnitude of each mode's contribution (determined from the fit) is shown as the orange, light blue, and green curves in the left column.

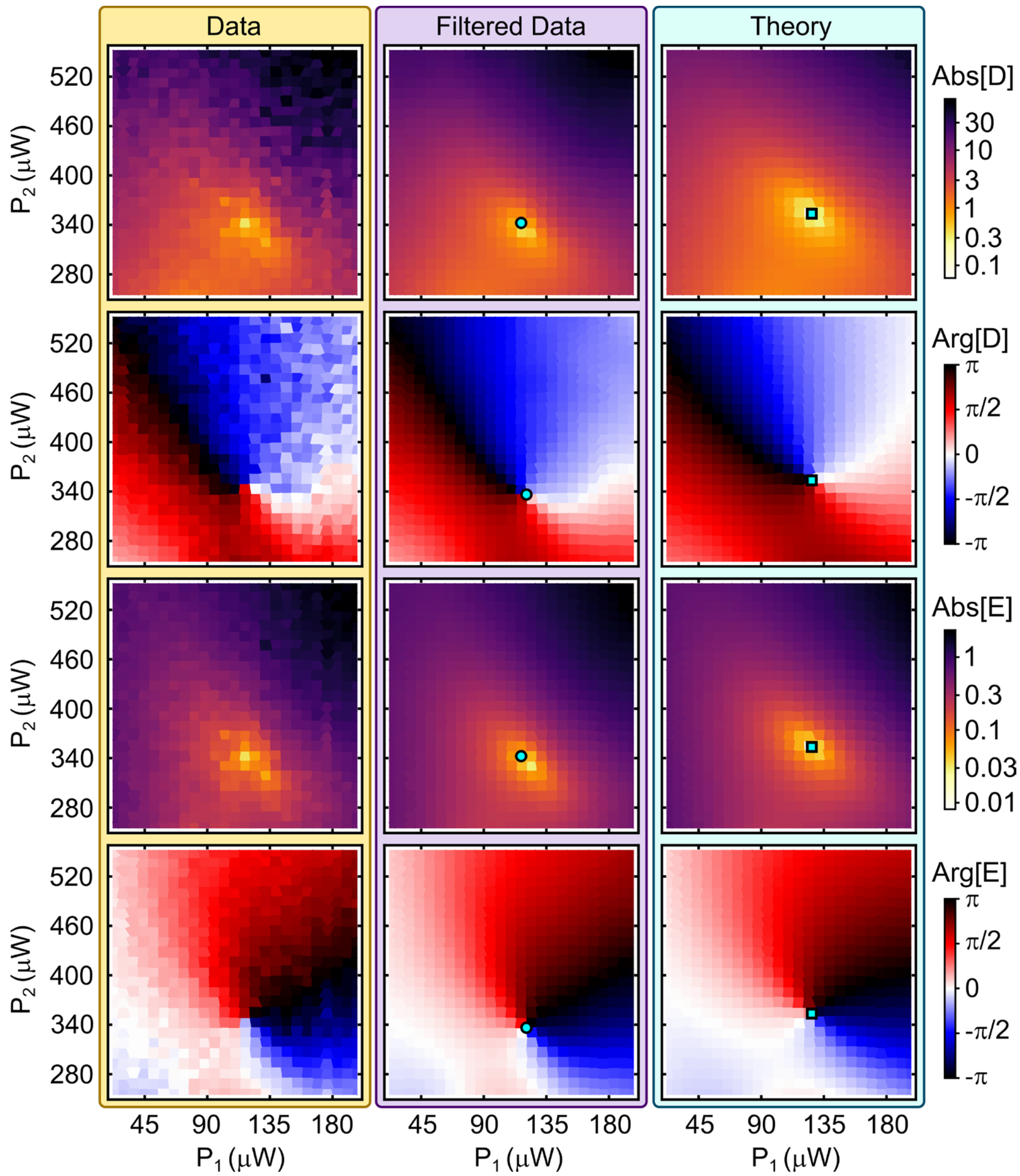

**Fig. 2 | Locating EP₂ points on the hypersurface $\mathcal{S}$.** The complex-valued quantities $D$ and $E$ measured on a typical 2D sheet in the hypersurface $\mathcal{S}$. The units of $D$ are $10^{10} \times (2\pi\ Hz)^6$. For this sheet, $P_3 = 78$ μW and $\delta = 2\pi \times 60$ kHz. Left column: raw data. Middle column: data after outlier rejection and smoothing (see Supplementary Information). Cyan circles: algorithmically identified $\boldsymbol{\Psi}_{\mathrm{EP2}}$. Right column: $D$ and $E$ calculated from optomechanics theory. Cyan squares: $\boldsymbol{\Psi}_{\mathrm{EP2}}$ determined from this calculation. Data from the other 60 sheets is shown in the Supplementary Information.

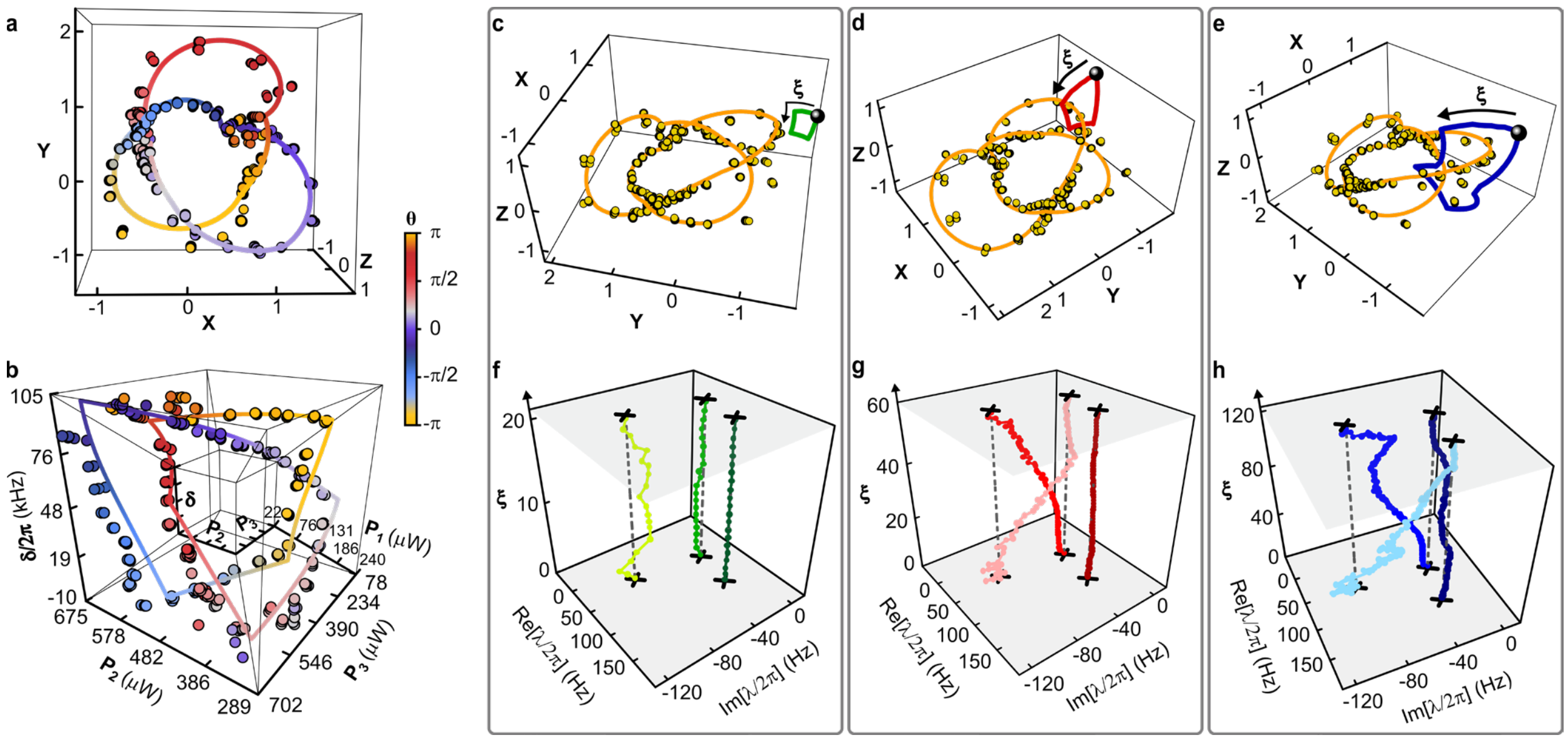

**Fig. 3 | Measurements of the EP₂ knot $\mathcal{K}$ and the eigenvalue braids. a,** All of the EP₂ locations ($\boldsymbol{\Psi}_{\mathrm{EP2}}$) shown in a stereographic projection of $\mathcal{S}$. $X$, $Y$, and $Z$ are dimensionless combinations of the control parameters. **b,** The data from **a** in a projection where each of the six "faces" of $\mathcal{S}$ that contains $\boldsymbol{\Psi}_{\mathrm{EP2}}$ is linearly mapped to one of the hexahedrons surrounding the central cube. The solid curve in **a**, **b** is the best fit to the data. Details of the projections and fit are in Methods. The coordinate $\theta$ is described in Supplementary Information. **c-e,** Three control loops (green, red, blue) in $\mathcal{S}$, each from a different $\ell$ and sharing a common basepoint (black sphere). The measured knot $\mathcal{K}$ (yellow circles) and the best-fit knot (orange curve) are shown for reference. The projection is the same as in **a**. **f-h,** The eigenvalue spectrum $\boldsymbol{\lambda}(\boldsymbol{\Psi})$ as $\boldsymbol{\Psi}$ is varied around the corresponding loop from **c-e**. $\xi$ indexes the values of $\boldsymbol{\Psi}$ along each loop. The black crosses show $\boldsymbol{\lambda}$ at the basepoint. The dashed lines are guides to the eye. The 1-$\sigma$ confidence intervals for $\boldsymbol{\lambda}$ are comparable to the size of the plotted points. The measured $\boldsymbol{\lambda}$ traces out the braids: $\mathbb{I}$ (**f**), $\sigma_1$ (**g**), and $\sigma_2\sigma_1$ (**h**). Extended Data Fig. 10 shows the control loops, and Extended Data Fig. 6 shows a comparison with theory. Animations of this figure are shown in Supplementary Videos 3 and 4.

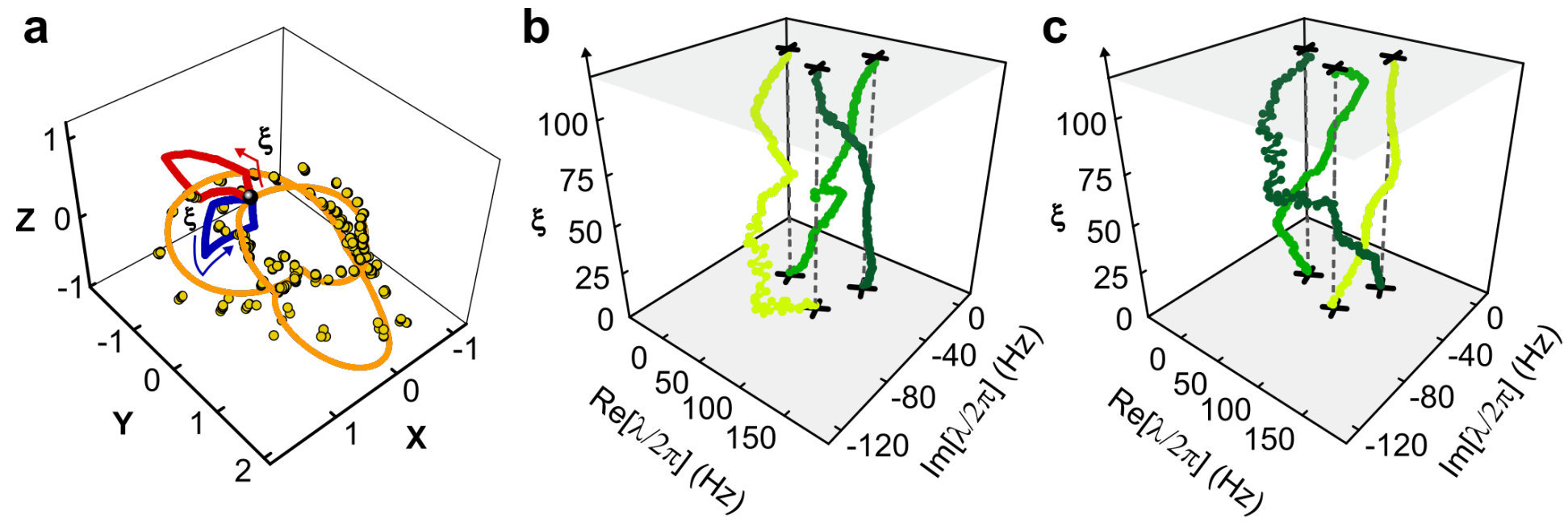

**Fig. 4 | Non-commutation of control loops. a,** Two loops (red, blue) belonging to different $\ell$. They are non-intersecting, except that they have a common basepoint (black sphere). The measured knot $\mathcal{K}$ (yellow circles) and the best-fit knot (orange curve) are shown for reference. The projection is the same as in Fig. 3a. **b,** The spectrum $\boldsymbol{\lambda}(\boldsymbol{\Psi})$ as $\boldsymbol{\Psi}$ is varied around the loop formed by concatenating the two loops in **a**. Specifically, the red loop is traversed first ($1 \leq \xi \leq 59$), and then the blue loop ($60 \leq \xi \leq 116$). The black crosses show $\boldsymbol{\lambda}$ at the basepoint. The dashed lines are guides to the eye, and the 1-$\sigma$ confidence intervals for $\boldsymbol{\lambda}$ are comparable to the size of the plotted points. **c,** The spectrum $\boldsymbol{\lambda}(\boldsymbol{\Psi})$ as $\boldsymbol{\Psi}$ is varied first around the blue loop ($1 \leq \xi \leq 57$), and then the red loop ($58 \leq \xi \leq 116$). In both cases, the loops are traversed in the sense indicated by the arrows in **a**.

# Methods

**Characteristic polynomial, discriminant, and the trefoil knot**

For an $N \times N$ matrix $\mathbf{H}$, the eigenvalues are the solutions of the characteristic equation $\det(\lambda\mathbf{I} - \mathbf{H}) = 0$, which can be written as

$$\lambda^N - a_1\lambda^{N-1} + a_2\lambda^{N-2} + \cdots + (-1)^N a_N = 0 .$$

The coefficients $a_i$ are invariants of $\mathbf{H}$ under similarity transforms (change of basis), and in particular $a_1 = \mathrm{tr}\,\mathbf{H}$ and $a_N = \det\mathbf{H}$. The characteristic polynomial on the left-hand side of this equation can be factored as $\prod_{i=1}^{N}(\lambda - \lambda_i)$ where the roots $\lambda_i$ may be repeated. The coefficients $a_i$ are the elementary symmetric polynomials in the roots $\lambda_i$, namely $a_1 = \sum_{i=1}^{N}\lambda_i$, $a_2 = \sum_{i,j:i<j}\lambda_i\lambda_j$, … , $a_N = \prod_{i=1}^{N}\lambda_i$.

The discriminant of the polynomial is defined as $D = \prod_{i<j}(\lambda_i - \lambda_j)^2$; it vanishes if any two roots of the polynomial are equal. Being a symmetric polynomial, it can be expressed algebraically in terms of the elementary symmetric polynomials $a_i$ (for example, Ref. [41]). The explicit expressions are simpler if we shift $\mathbf{H}$ by a multiple of the identity so that $a_1 = \mathrm{tr}\,\mathbf{H} = 0$. Then for the quadratic, $N = 2$, the discriminant is the familiar expression $D = -4a_2$, and the roots are $\pm\sqrt{-a_2}$. Focusing hereafter on the cubic, $N = 3$, its discriminant is[41]

$$D = -4a_2^3 - 27a_3^2 ,$$

which enters the formulas for the roots.

Defining $x = a_3$, $y = -a_2$ as our coordinates in $\mathbb{C}^2$ (so that the characteristic polynomial is $p_{\mathbf{H}} = \lambda^3 - y\lambda - x$), the solutions to the equation $D(x,y) = 4y^3 - 27x^2 = 0$ form an algebraic variety (a hypersurface) in $\mathbb{C}^2$ which has a singularity at $x = y = 0$. $D$ has a (weighted) scaling property, such that rescaling $x \to ax$, $y \to by$, where $a, b$ are real and positive and $a^2 = b^3$, changes $D$ by a factor; this maps any nonzero solution to $D = 0$ to another. Thus the variety resembles a cone in $\mathbb{C}^2 \cong \mathbb{R}^4$, and it suffices to consider a cross section, that is the intersection of the variety with a hypersurface that (i) has the topology of a hypersphere $\mathcal{S}^3$, (ii) surrounds the origin without passing through it or intersecting itself, and (iii) is everywhere transverse to the local action of an infinitesimal (say, by $a = 1 + \varepsilon$, $\varepsilon$ small) such scaling. Any two such hypersurfaces are isotopy equivalent (via a rescaling that depends on position on the hypersurface). A particular such hypersurface[42] results from considering the unit hypersphere defined by $|x|^2 + |y|^2 = 1$. The points $(x,y)$ on the hypersphere that satisfy $D(x,y) = 0$ can be parameterized as $(x,y) = (r_x e^{3i\theta}, r_y e^{2i\theta})$, where $r_x, r_y$ are real positive constants, and $0 \le \theta < 2\pi$ is a variable. These points lie on a 2-torus $T^2$ embedded in $\mathcal{S}^3$, and form a closed curve which is a trefoil knot in this $\mathcal{S}^3$.[42] This is illustrated in Extended Data Fig. 1, as described in the Supplementary Information.

For any knot or link in $\mathbb{R}^3$ or $\mathcal{S}^3$, the fundamental group of its complement is an isotopy invariant of the knot or link called the knot group. The knot group of the trefoil is well-known to be the braid group $B_3$, or this can be inferred by reasoning as in the main text.

**Experimental setup**

As described in the main text, this experiment focuses on three vibrational modes of a $Si_3N_4$ membrane. The membrane's dimensions are 1 mm × 1 mm × 50 nm. These modes' bare eigenvalues (i.e., in the absence of optomechanical effects) are $\tilde{\boldsymbol{\lambda}}^{(0)} = (\tilde{\lambda}_1^{(0)}, \tilde{\lambda}_2^{(0)}, \tilde{\lambda}_3^{(0)}) = 2\pi \times (352\,243 - 2.2i, 557\,217 - 1.9i, 704\,837 - 1.8i)$ Hz, where the real (imaginary) parts give each mode's oscillation frequency (amplitude damping rate). Frequencies related to the mechanical modes are denoted with a tilde when given in the lab frame, and without a tilde in the frame $\mathcal{R}$ described below.

The dynamical matrix $\widetilde{\mathbf{H}}$ governing these modes is controlled using the dynamical back-action (DBA) effect of cavity optomechanics.[31] The membrane is placed in an optical cavity with linewidth $\kappa/2\pi = 190$ kHz, input coupling rate $\kappa_{\text{in}} = 0.267\,\kappa$, and optomechanical coupling rates $\boldsymbol{g} = (g_1, g_2, g_3) = 2\pi \times (0.198, 0.304, 0.300)$ Hz. Additional details of the apparatus are in the Supplementary Information, Extended Data Fig. 8, and Ref. [43].

The cavity is driven with three tones produced from a single laser ("control", Fig. 1a). The DBA from each tone induces a complex-valued shift in each mechanical mode's eigenvalue.[31] In addition, each pair of tones gives rise to an intracavity beatnote, which induces a complex-valued coupling between pairs of modes whose frequency difference is comparable to the beatnote frequency.[44,45] In the resolved sideband regime ($\kappa \ll \tilde{\omega}_{1,2,3}^{(0)}$, where $\tilde{\omega}_i^{(0)} \equiv \text{Re}(\tilde{\lambda}_i^{(0)})$) these shifts and couplings can be tuned over the complex plane by varying the tones' powers $P_k$ and detunings $\Delta_k$ ($k \in \{1,2,3\}$). An expression for $\widetilde{\mathbf{H}}$ in terms of $P_k$, $\Delta_k$, $\kappa$, $\kappa_{\text{in}}$, $\boldsymbol{g}$, and $\tilde{\boldsymbol{\lambda}}^{(0)}$ is given in the Supplementary Information. For the experiments described here, the tones' common detuning $\delta$ (Fig. 1b) is varied. Their relative detunings are fixed, and are chosen to produce beatnote frequencies close to the differences between the $\tilde{\omega}_i^{(0)}$.

The beatnote frequencies are chosen so that there is a rotating frame $\mathcal{R}$ (defined in the Supplementary Information) in which the dynamical matrix $\mathbf{H}$ is time-independent (in the rotating wave approximation), and in which the bare eigenvalues $\boldsymbol{\lambda}^{(0)}$ are almost degenerate (their non-degeneracy in $\mathcal{R}$ is set by $\eta = -2\pi \times 100$ Hz (Fig. 1b)).

Thus, within $\mathcal{R}$ the mechanical modes can be described by the equation of motion

$$\dot{\boldsymbol{x}}(t) = -i\mathbf{H}(\boldsymbol{\Psi})\boldsymbol{x}(t) + \boldsymbol{f}(t)$$

Here $\boldsymbol{x}(t) = (x_1(t), x_2(t), x_3(t))^T$ and $\boldsymbol{f}(t) = (f_1(t), f_2(t), f_3(t))^T$ are the modes' complex-valued amplitudes and the external forces driving them. While the above equation is the generic equation of motion for any linear system, we emphasize the form of $\mathbf{H}(\boldsymbol{\Psi})$ realized here:

specifically, that the controls $\boldsymbol{\Psi}$ completely and smoothly parametrize all of the complex eigenspectra in a neighborhood that includes $EP_3$.[6]

**Locating the $EP_3$**

This section gives the protocol for experimentally identifying the $EP_3$. Approaches to identifying three-fold degeneracies are also given in Refs. [16,17,18,19,20,21,22,23,24,46,47,48].

We identify the value of control parameters ($\boldsymbol{\Psi}_{\mathrm{EP3}}$) that corresponds to $EP_3$ through the quantity $d = |\lambda_1 - \lambda_2| + |\lambda_2 - \lambda_3| + |\lambda_3 - \lambda_1|$, which may be visualized as the perimeter of the triangle formed by the system's three eigenvalues $\boldsymbol{\lambda}$ in the complex plane. At $\boldsymbol{\Psi}_{\mathrm{EP3}}$ the three eigenvalues are equal, and so $d = 0$.

The search for the $EP_3$ point starts with the estimate:

$$\boldsymbol{\Psi}_{\mathrm{EP3}}^{(\mathrm{thy})} = (2\pi \times 49.7\ \mathrm{kHz}, 115\ \mathrm{\mu W}, 387\ \mathrm{\mu W}, 285\ \mathrm{\mu W})\ .$$

We proceed by fixing three of the control parameters to these values, and scanning the fourth (say, $\Psi_i$). At each value of $\boldsymbol{\Psi}$ in this 1D sweep, $\boldsymbol{\lambda}$ is measured (as described in the Supplementary Information) and converted to $d(\boldsymbol{\Psi})$. The experimental estimate $\boldsymbol{\Psi}_{\mathrm{EP3}}^{(\mathrm{est})}$ is then updated with the value of $\Psi_i$ that minimizes $d$ over that sweep. This process is then repeated for different choices of $\Psi_i$. The estimate resulting from these 1D sweeps is:

$$\boldsymbol{\Psi}_{\mathrm{EP3}}^{(\mathrm{est})} = (2\pi \times 49.7\ \mathrm{kHz}, 125\ \mathrm{\mu W}, 435\ \mathrm{\mu W}, 300\ \mathrm{\mu W})\ .$$

To further refine the estimate of $\boldsymbol{\Psi}_{\mathrm{EP3}}$, we then measure $d(\boldsymbol{\Psi})$ on 2D sheets that pass through $\boldsymbol{\Psi}_{\mathrm{EP3}}^{(\mathrm{est})}$. For each 2D sheet, two control parameters are scanned while the other two are fixed, resulting in a total of six sheets. The $d(\boldsymbol{\Psi})$ measured on these sheets are shown in Extended Data Fig. 2. For visualization purposes, Extended Data Fig. 3 shows the same sheets, but arranged in 3D to illustrate that $d(\boldsymbol{\Psi})$ is minimized in the neighborhood of $\boldsymbol{\Psi}_{\mathrm{EP3}}^{(\mathrm{est})}$. In Extended Data Figs. 2,3, the middle row shows the filtered data (see the Supplementary Information for details of the filtering), and the bottom row shows the values of $d(\boldsymbol{\Psi})$ calculated from $\mathbf{H}$ (Supplementary Information) using the best-fit optomechanical parameters determined by fitting the knot, as shown in Figs. 3a,b and described in detail in the Supplementary Information.

Near to $\boldsymbol{\Psi}_{\mathrm{EP3}}$ the quantity $d(\boldsymbol{\Psi})$ is expected to scale as[4] $d(\boldsymbol{\Psi}) \sim |\boldsymbol{\Psi} - \boldsymbol{\Psi}_{\mathrm{EP3}}|^{1/3}$ , but in practice the sharp cusp in $d(\boldsymbol{\Psi})$ is broadened by fluctuations in $\boldsymbol{\Psi}$. Nevertheless, clear minima are visible in the measured $d(\boldsymbol{\Psi})$, and their locations are given in the Supplementary Information (see Table 1). Details of the algorithm used to identify the minima are also in the Supplementary Information. The mean location of these minima is taken as the experimentally identified $EP_3$:

$$\boldsymbol{\Psi}_{\mathrm{EP3}}^{(\mathrm{exp})} = \left(\delta_{\mathrm{EP3}}^{(\mathrm{exp})}, P_{1,\mathrm{EP3}}^{(\mathrm{exp})}, P_{2,\mathrm{EP3}}^{(\mathrm{exp})}, P_{3,\mathrm{EP3}}^{(\mathrm{exp})}\right)$$

$$= (2\pi \times 54(7)\ \text{kHz}, 128(8)\ \mu\text{W}, 428(3)\ \mu\text{W}, 304(15)\ \mu\text{W})$$

This compares well with the location of $EP_3$ that is obtained from the best-fit parameters returned by fitting the measured knot:

$$\boldsymbol{\Psi}_{\text{EP3}}^{(\text{knot})} = (2\pi \times 60.2\ \text{kHz}, 116\ \mu\text{W}, 477\ \mu\text{W}, 329\ \mu\text{W}).$$

**Projections of the hypersurface $\mathcal{S}$**

Here we describe the two projections that are used in Figs. 3a,b of the main text (as well as Extended Data Fig. 4) to represent data acquired on the hypersurface $\mathcal{S}$, which is the surface of a 4D hyperrectangle.

***Standard stereographic projection:*** Stereographic projection is a standard means for representing a sphere (typically of 1, 2 or 3 dimensions) in a Euclidean space with the same number of dimensions. In Fig. 3a of the main paper, we represent $\mathcal{S}$ by first projecting it onto the unit 3-sphere $\mathcal{S}^3$ and then applying the standard stereographic projection of $\mathcal{S}^3$ onto $\mathbb{R}^3$.

The map is constructed by first adimensionalizing the control parameter as

$$\boldsymbol{\Psi}' \coloneqq \frac{\boldsymbol{\Psi}}{\boldsymbol{\Psi}_{\text{EP3}}^{(\text{exp})}} - 1 \coloneqq \left(\frac{\delta}{\delta_{\text{EP3}}^{(\text{exp})}} - 1, \frac{P_1}{P_{1,\text{EP3}}^{(\text{exp})}} - 1, \frac{P_2}{P_{2,\text{EP3}}^{(\text{exp})}} - 1, \frac{P_3}{P_{3,\text{EP3}}^{(\text{exp})}} - 1\right)$$

and then normalizing it as $\boldsymbol{\Psi}'' \coloneqq \frac{\boldsymbol{\Psi}'}{\|\boldsymbol{\Psi}'\|}$, where $\|\cdot\|$ is the conventional $L^2$ norm. Note here the implicit use of the fact that $\boldsymbol{\Psi}_{\text{EP3}}^{(\text{exp})}$ lies inside the 4-volume bounded by $\mathcal{S}$.

Next, we act on $\boldsymbol{\Psi}''$ with a 4D rotation $\mathbf{R}$ (specified below). The new unit vector $\mathbf{R}\boldsymbol{\Psi}'' \equiv (x, z, w, y)$ is then stereographically projected onto the 3D cartesian coordinates $(X, Y, Z)$ as $X = \frac{x}{1-w}, Y = \frac{y}{1-w}, Z = \frac{z}{1-w}$. Thus, the choice of $\mathbf{R}$ corresponds to choosing the pole $(x, z, w, y) = (0,0,1,0)$ for the stereographic projection.

The same pole is chosen for all the stereographic projections shown in this work (except for Extended Data Fig. 1, whose generation is described in the Supplementary Material). It is chosen so as to optimize the visualization of the experimentally identified knot, and corresponds to $\boldsymbol{\Psi}'' = (0.1, -0.83, 0.55, 0)$, or equivalently, $\boldsymbol{\Psi} = (2\pi \times 55\ \text{kHz}, 22\ \mu\text{W}, 596\ \mu\text{W}, 304\ \mu\text{W})$.

***"Rectilinear stereographic" projection:*** The projection shown in Fig. 3b of the main text is isomorphic to the projection just described. However, it is intended to provide a more intuitive representation of the dimensionful experimental parameters $\boldsymbol{\Psi}$. Animations that describe this projection are shown in Videos 1 & 2 (Supplementary Information).

This projection consists of five steps.

(i) We select one of the eight 3D hyperrectangles that constitute $\mathcal{S}$

(ii) We simply rescale its axes so that it forms a cube (this is the central cube in Fig. 3b)

(iii) Each of the six 3D hyperrectangles adjacent to the first one is also rescaled to form a cube, which is then attached to the first cube on their common 2D face. The resulting "six-way cross" faithfully represents the connections of the central cube to its six neighbors

(iv) To faithfully represent the connections among these six neighbors, a bilinear transformation[43] is applied to each, deforming each cube into a truncated square pyramid. The transformation is chosen so that the 2D faces that are common to any two of these neighbors are made to touch. These seven hexahedrons (the central cube and the six truncated square pyramids surrounding it) can readily be labelled by their original dimensionful axes, as in Fig. 3b. Together they form the bounding box of Fig. 3b.

(v) The final 3D hyperrectangle is mapped to the exterior of the bounding box via a nonlinear mapping, and extends to infinity (as does the standard stereographic projection described above).

As described in the Supplementary Information, there are no $EP_2$'s in the two 3D hyperrectangles that correspond to constant $P_1$. We choose these to be the interior (cubical) hexahedron and the exterior region. This choice places all of the $EP_2$'s in the six truncated square pyramids, facilitating a clear view of the knot. Supplementary Video 2 gives animated views of the data and fit shown in this projection.

**Fitting the $\mathbf{EP_2}$ locations to the optomechanical model**

This section describes the fit of the three-mode optomechanical model to the 291 experimentally identified $EP_2$ points shown in Figs. 3a,b of the main text. These locations are denoted as $\boldsymbol{\Psi}_{\mathrm{EP2}}^{(\exp,\ell)}$, with $1 \leq \ell \leq 291$.

The best-fit parameters $\boldsymbol{g}$ and $\kappa$ for the model are obtained by minimizing the cost function

$$C(\boldsymbol{g},\kappa) = \sum_\ell \left| \boldsymbol{\Psi}_{\mathrm{EP2}}^{(\exp,\ell)} - \boldsymbol{\Psi}_{\mathrm{EP2}}^{(\mathrm{thy},\ell)}(\boldsymbol{g},\kappa) \right|^2$$

where the summands define a distance between the experiment and theory, which is adimensionalized by the $EP_3$ coordinates $\boldsymbol{\Psi}_{\mathrm{EP3}}^{(\exp)} = \left(\delta_{\mathrm{EP3}}^{(\exp)}, P_{1,\mathrm{EP3}}^{(\exp)}, P_{2,\mathrm{EP3}}^{(\exp)}, P_{3,\mathrm{EP3}}^{(\exp)}\right)$.

In particular, for

$$\boldsymbol{\Psi}_{\mathrm{EP2}}^{(\exp,\ell)} = \left(\delta_{\mathrm{EP2}}^{(\exp,\ell)}, P_{1,\mathrm{EP2}}^{(\exp,\ell)}, P_{2,\mathrm{EP2}}^{(\exp,\ell)}, P_{3,\mathrm{EP2}}^{(\exp,\ell)}\right)$$

and

$$\boldsymbol{\Psi}_{\mathrm{EP2}}^{(\mathrm{thy},\ell)} = \left(\delta_{\mathrm{EP2}}^{(\mathrm{thy},\ell)}, P_{1,\mathrm{EP2}}^{(\mathrm{thy},\ell)}, P_{2,\mathrm{EP2}}^{(\mathrm{thy},\ell)}, P_{3,\mathrm{EP2}}^{(\mathrm{thy},\ell)}\right)$$

this dimensionless distance (squared) is

$$\left|\boldsymbol{\Psi}_{\mathrm{EP2}}^{(\exp,\ell)} - \boldsymbol{\Psi}_{\mathrm{EP2}}^{(\mathrm{thy},\ell)}(\boldsymbol{g},\kappa)\right|^2 := \left(\frac{\delta_{\mathrm{EP2}}^{(\exp,\ell)}-\delta_{\mathrm{EP2}}^{(\mathrm{thy},\ell)}}{\delta_{\mathrm{EP3}}^{(\exp)}}\right)^2 + \left(\frac{P_{1,\mathrm{EP2}}^{(\exp,\ell)}-P_{1,\mathrm{EP2}}^{(\mathrm{thy},\ell)}}{P_{1,\mathrm{EP3}}^{(\exp)}}\right)^2 + \left(\frac{P_{2,\mathrm{EP2}}^{(\exp,\ell)}-P_{2,\mathrm{EP2}}^{(\mathrm{thy},\ell)}}{P_{2,\mathrm{EP3}}^{(\exp)}}\right)^2 + \left(\frac{P_{3,\mathrm{EP2}}^{(\exp,\ell)}-P_{3,\mathrm{EP2}}^{(\mathrm{thy},\ell)}}{P_{3,\mathrm{EP3}}^{(\exp)}}\right)^2$$

Here, $\boldsymbol{\Psi}_{\mathrm{EP2}}^{(\mathrm{thy},\ell)}(\boldsymbol{g},\kappa)$ is the $\mathrm{EP}_2$ point found numerically (as a root of the discriminant $D(\boldsymbol{\Psi},\boldsymbol{g},\kappa)$, see the Supplementary Information) in a neighborhood of $\boldsymbol{\Psi}_{\mathrm{EP2}}^{(\exp,\ell)}$ and within its 2D data sheet. For example, if $\boldsymbol{\Psi}_{\mathrm{EP2}}^{(\exp,\ell)}$ is identified in a data sheet that rasters $P_1$ and $P_2$ while holding $\delta$ and $P_3$ fixed, the numerical root is found within the neighborhood

$$\left(0.65\, P_{1,\mathrm{EP2}}^{(\exp,\ell)}, 1.35\, P_{1,\mathrm{EP2}}^{(\exp,\ell)}\right) \times \left(0.65\, P_{2,\mathrm{EP2}}^{(\exp,\ell)}, 1.35\, P_{2,\mathrm{EP2}}^{(\exp,\ell)}\right)$$

at the same fixed values of $\delta$ and $P_3$. $\boldsymbol{\Psi}_{\mathrm{EP2}}^{(\mathrm{thy},\ell)}(\boldsymbol{g},\kappa)$ is evaluated with $\kappa_{\mathrm{in}}/\kappa = 0.267$, and $\tilde{\boldsymbol{\lambda}}^{(0)}$ held equal to the values determined from the single-tone DBA measurements described in the Supplementary Information.

The minimization of $C(\boldsymbol{g},\kappa)$ is implemented numerically on a high-performance cluster. The best-fit parameters so obtained are:

$$\boldsymbol{g} = 2\pi \times (0.1979, 0.3442, 0.3092)\ \mathrm{Hz}$$
$$\kappa = 2\pi \times 173.84\ \mathrm{kHz}$$

These parameters are used to produce the "best-fit knot" shown as the continuous curve in Figs. 3a,b in the main text. This curve is generated by using the best-fit values of $\boldsymbol{g}$ and $\kappa$ given just above to calculate $\boldsymbol{\lambda}$ on 16,000 2D sheets in $\mathcal{S}$. On each sheet, the $\mathrm{EP}_2$ points are identified as the roots of the discriminant $D$ (found numerically as described in the Supplementary Information). At each of these $\mathrm{EP}_2$ points, $\theta$ is also calculated. Finally, these points are colored according to $\theta$ and are connected by straight line segments.

The values of the parameters $\boldsymbol{g}$ and $\kappa$ given just above are also used to generate the theory plots in Fig. 2 of the main text, and in Extended Data Figs. 2,3,5 and 6 and Supplementary Video 5.

**Relation of the present work to studies of non-Hermitian band structure**

Topics related to those described in this work have been considered in the context of non-Hermitian band structure (NHBS).[49,50,22,51,52,53,54,55,56,57,58,59] However there are a number of qualitative differences between NHBS and the non-Hermitian oscillators considered here: in the physical systems being described, the mathematical concepts relevant to the description, and the genericness of the resulting topological structures.

The physical system under consideration in NHBS is a wave propagating in an $L$-dimensional lattice (where $L$ is typically 1, 2, or 3) that possesses a combination of non-reciprocity, gain, and loss. Propagation in such a lattice can be characterized by bands whose dispersion is given by the complex eigenvalues of a matrix (which plays the role of $\mathbf{H}$ in the present paper). A central question in NHBS is how these eigenvalues depend upon the quasimomentum $\mathbf{k}$ (whose vector components play the role of the control parameters considered in the present paper). Theoretical[51,56,57] and experimental[22,58] work has shown that varying $\mathbf{k}$ in a closed loop may result in non-trivial eigenvalue braids. Theoretical work has shown that, for some lattices with $L = 3$, two-band systems described by $2 \times 2$ matrices may exhibit a trefoil knot of two-fold degeneracies within the Brillouin zone.[53,54,59] However we emphasize that these results are distinct from those presented here.

This is because in NHBS, the number of control parameters is limited to $L$, and the control space they span (the analog of $\mathcal{L}_N$ in the present work) is topologically non-trivial by assumption (since the Brillouin zone is an $L$-torus). In contrast, for non-Hermitian oscillators the control space ($\mathcal{L}_N$) is topologically trivial, and the number of controls (i.e., the dimensionality of $\mathcal{L}_N$) is sufficient to span the full space of complex eigenspectra. This results in the direct connection – described in the main text – between non-Hermitian oscillators and general complex polynomials. In particular, the non-trivial structure of the degenerate subspace (which establishes the correspondence between control loops and the non-Abelian group of eigenvalue braids) is a generic feature of $N \times N$ matrices. This is in contrast with NHBS, in which these features are not generic, but only appear upon fine tuning.

Lastly, we note that experiments on NHBS to date[22,58] have been limited to braids realized by two eigenvalues. Thus, they correspond to the $N = 2$ case, for which the subspace of degeneracies has a trivial geometry, and the group formed by the eigenvalue braids is Abelian. In contrast, for the $N = 3$ case explored in the experiments described here, the subspace of degeneracies has a non-trivial geometry, and the eigenvalue braids form a non-Abelian group.

Another approach to studying the propagation of linear excitations in non-Hermitian lattices is provided by gyroscopic metamaterials.[60,61] However, these systems possess purely real eigenvalues (because of the symplectic symmetry of their dynamical matrix), and so do not exhibit the behavior considered in this work.

# Methods references

**Acknowledgments**
This work was supported by Air Force Office of Scientific Research award no. FA9550-15-1-0270, Vannevar Bush Faculty Fellowship no. N00014-20-1-2628, and National Science Foundation grant no. DMR-1724923. We thank Y. Wang for helpful discussions, and the Yale Center for Research Computing for guidance and use of the research computing infrastructure, specifically M. Guy. J.G.E.H. thanks H. Vanderbilt. J.H. is now funded by Howard Hughes Medical Institute Janelia.

**Author Contributions**
N.R. and J.G.E.H. conceived the project. Y.S.S.P., J.H., P.A.H., C.G., L.J. and N.K. designed the experiment. Y.S.S.P., P.A.H. and C.G. took the data. Y.S.S.P., J.H., P.A.H., C.G, and Y.Z. analyzed and modelled the data. J.G.E.H. supervised the project. All authors contributed to the writing of the paper.

**Competing Interests**
The authors declare no competing interests.

**Data Availability**
The experimental data and numerical calculations are available from the corresponding authors upon reasonable request.

**Code Availability**
The code used for data analysis is available from the corresponding author upon reasonable request.

# Extended Data Figures

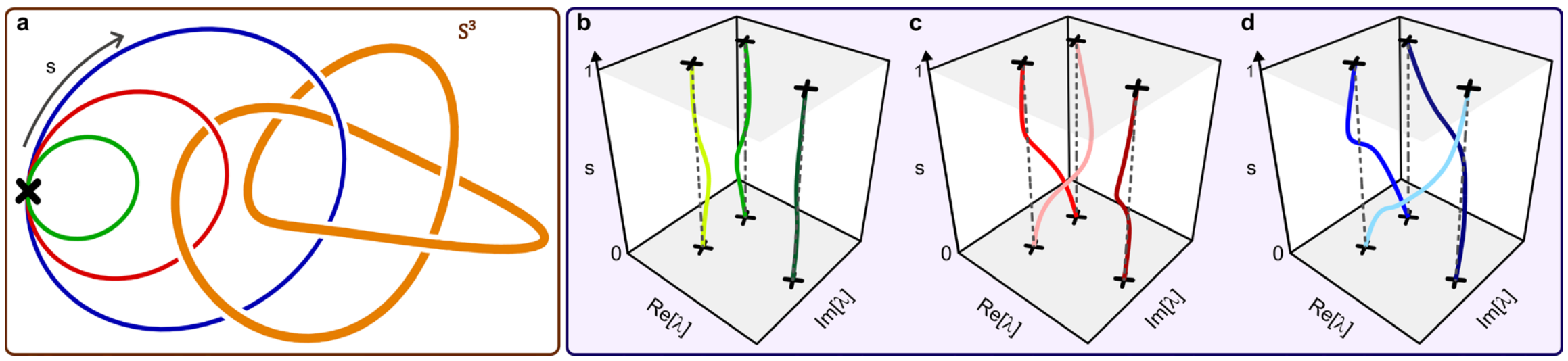

**Extended Data Fig. 1 | The trefoil knot of degeneracies and the eigenvalue braids for a three-mode system.** **a,** At a fixed distance from the three-fold degeneracy, the control space for the spectrum is $\mathcal{S}^3$ (shown here in stereographic projection). The degeneracies in this space are all two-fold and form a trefoil knot (orange). Three control loops (green, red, blue), each parameterized by $0 \leq s \leq 1$ share a common basepoint (black cross). **b-d,** Evolution of the eigenvalues as $s$ is varied around each loop in **a**. The black crosses show $\boldsymbol{\lambda}$ at the basepoint. The dashed lines are guides to the eye. This figure is calculated from the characteristic polynomial of a three-mode system (see the Supplementary Information)

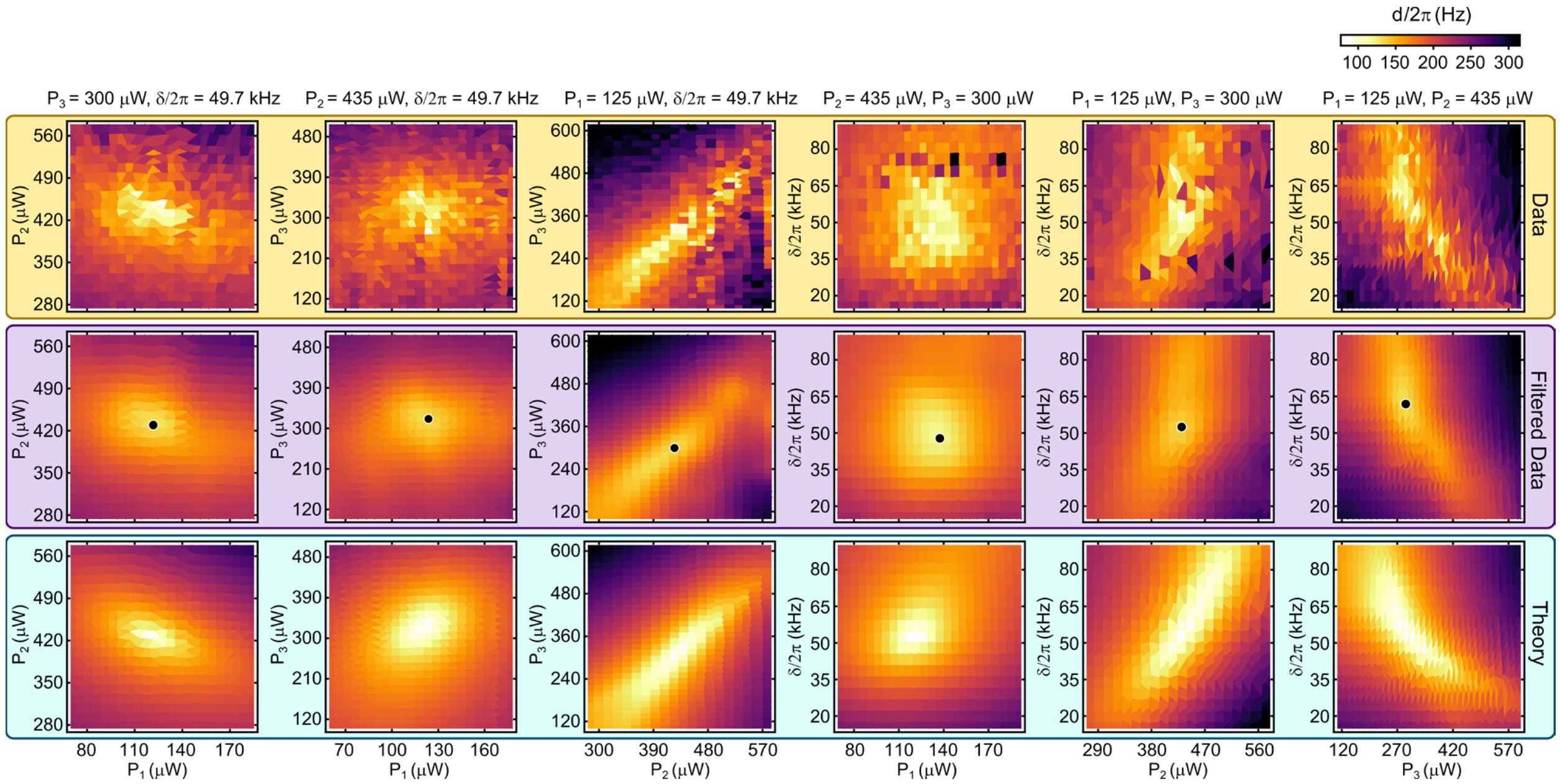

**Extended Data Fig. 2 | Locating EP₃.** The quantity $d(\boldsymbol{\Psi})$ (which ideally vanishes at $\boldsymbol{\Psi}_{\mathrm{EP3}}$), measured on six 2D sheets passing through $\boldsymbol{\Psi}_{\mathrm{EP3}}^{(\mathrm{est})}$, the location of the $\mathrm{EP}_3$ that is estimated from scanning individual components of $\boldsymbol{\Psi}$ (Methods). Top row: raw data. Middle row: data after outlier rejection and smoothing described in the Supplementary Information. The black circles show the minima that are located using the algorithm described in the Supplementary Information. Bottom row: the values of $d$ calculated from the optomechanical model.

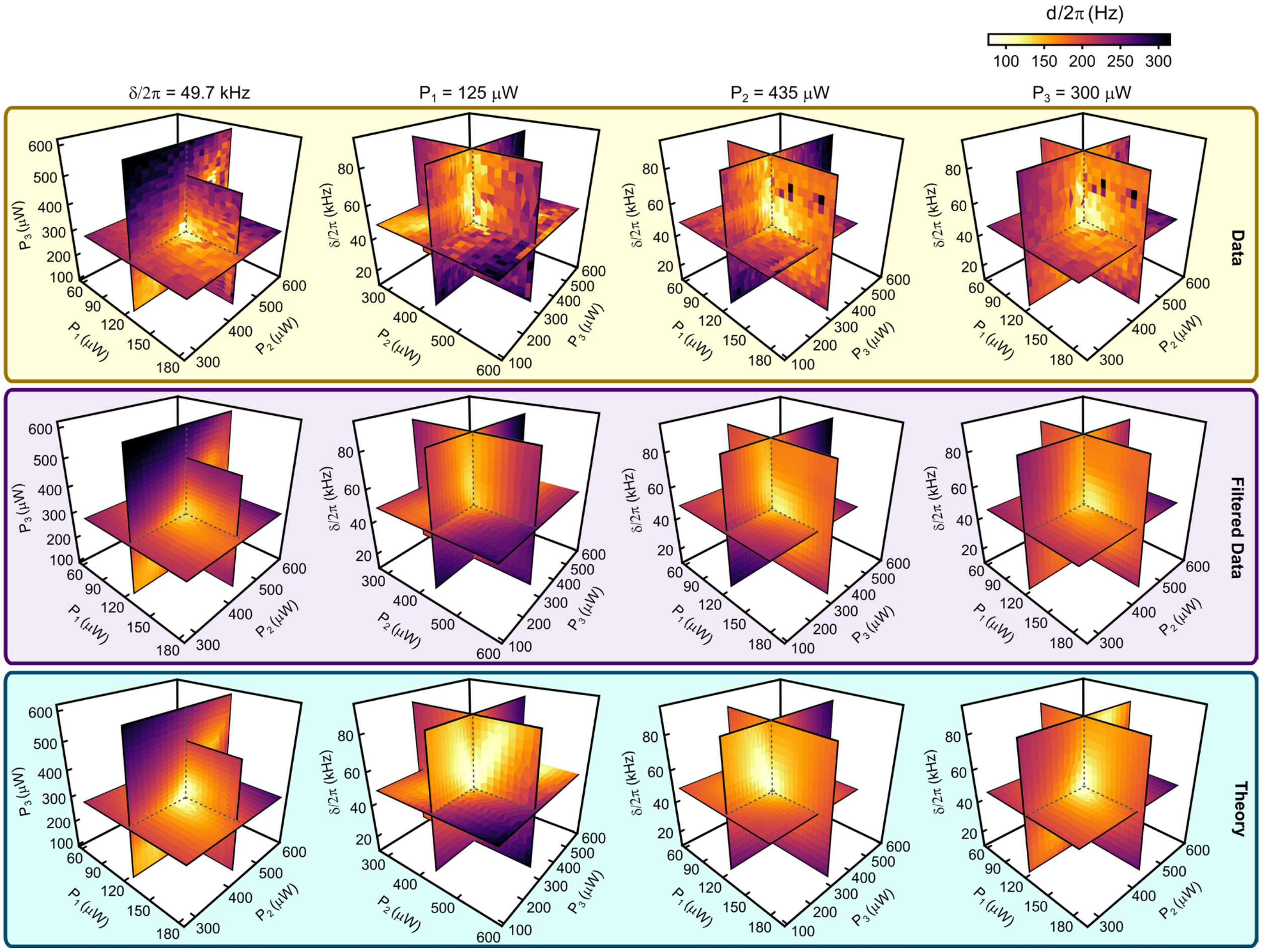

**Extended Data Fig. 3 | Locating EP₃ (perspective view).** The data of Extended Data Fig. 2 arranged in 3D to illustrate the minimum of $d(\boldsymbol{\Psi})$ in the neighborhood of the experimentally estimated location of the $EP_3$.

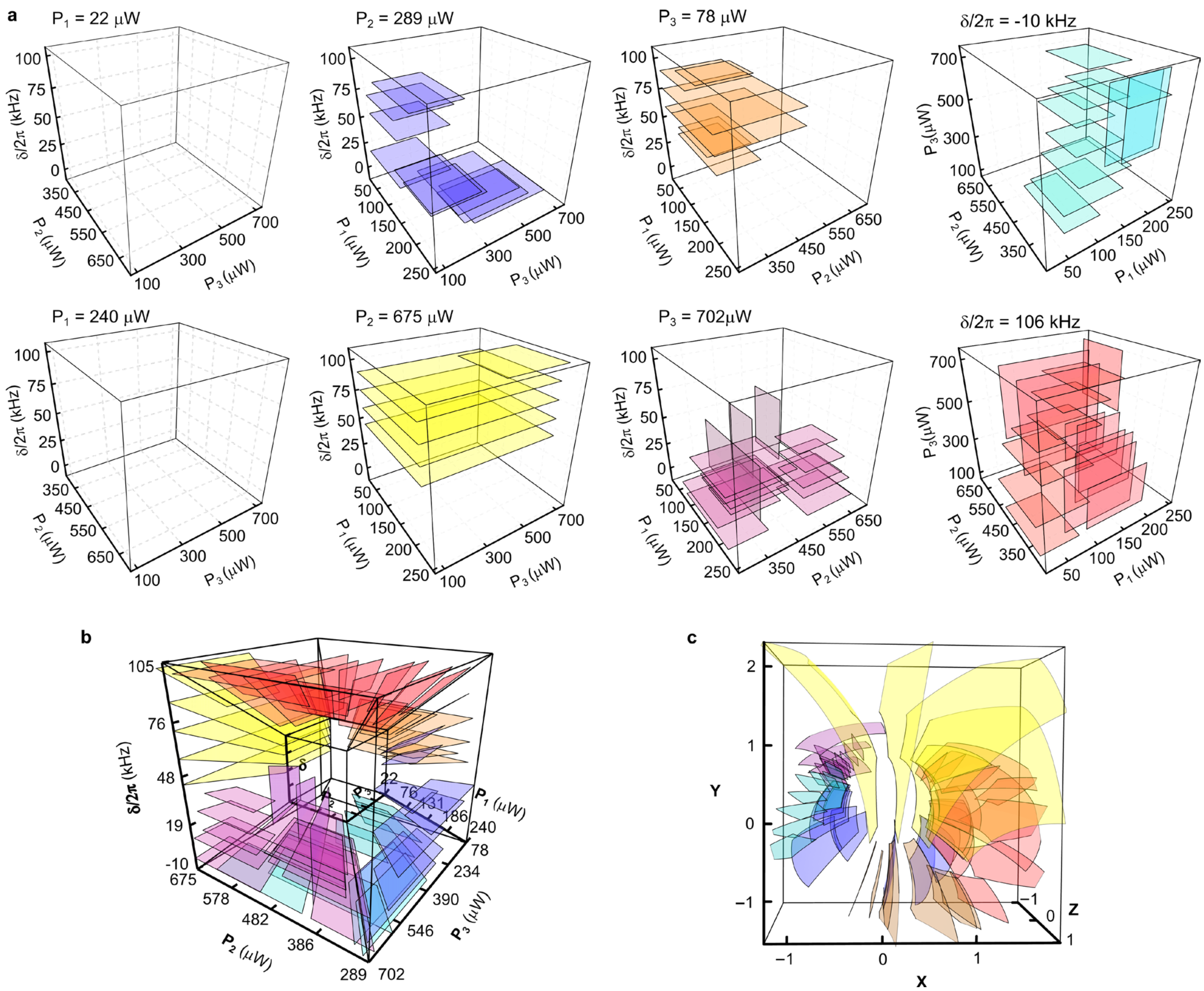

**Extended Data Fig. 4 | The locations of the sixty-one 2D sheets within $S$.** The sheets are color-coded by the 3D face in which they lie. **a,** The sheets are shown within each of the eight 3D faces of $S$. **b,** The same sheets as in **a**, shown using the "rectilinear stereographic" projection of Fig. 3b. Note that in this projection, all of the sheets are contained within the plot's bounding box. **c,** The same sheets, shown using the stereographic projection of Fig. 3a. The thin black lines show the boundary of each sheet. Thin gray lines show where a sheet exits the plot's bounding box. The projections are described in Methods. The data from these sheets is shown in Video 5 of the Supplementary Information.

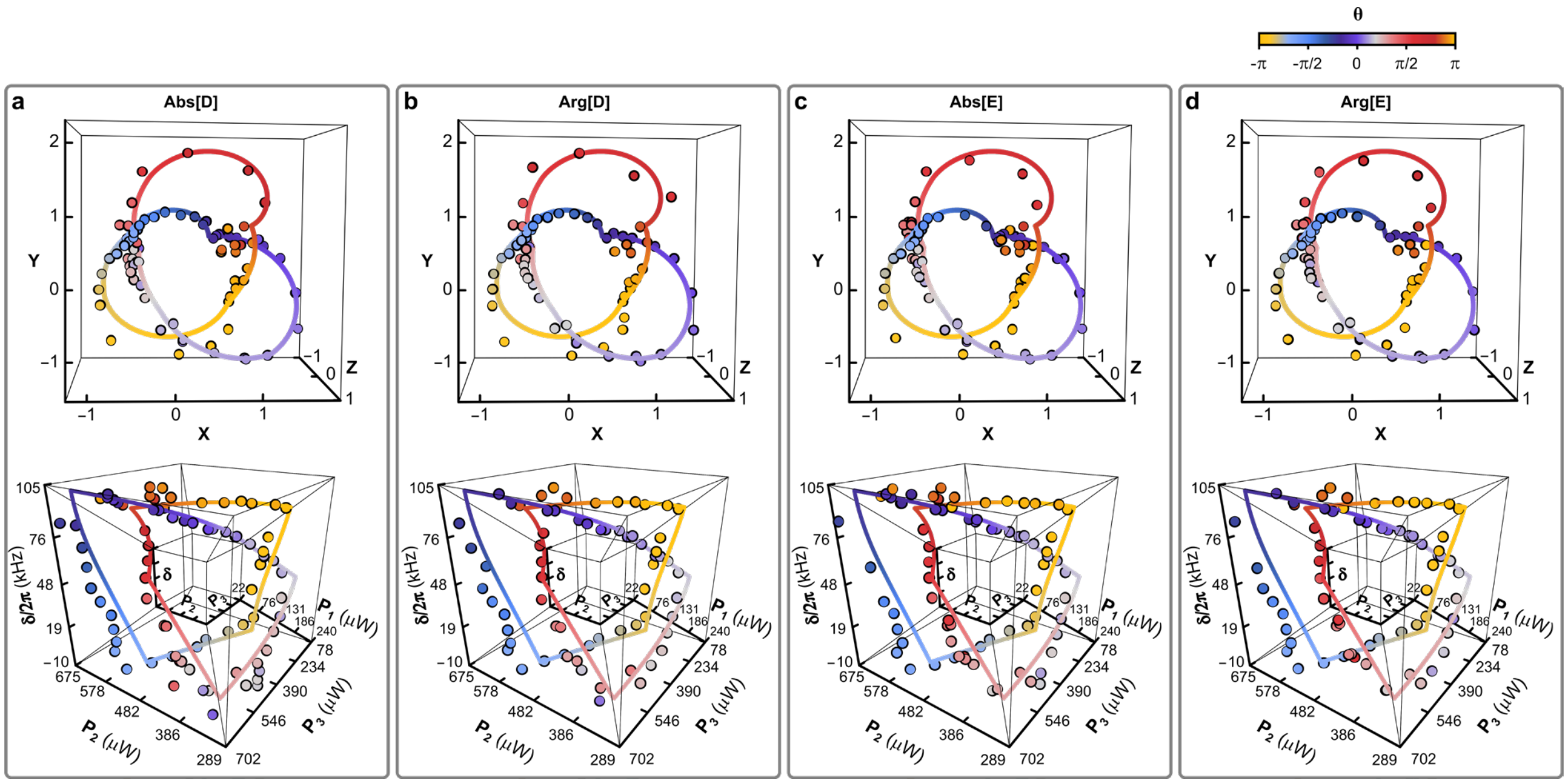

**Extended Data Fig. 5 | The knot of $EP_2$ via four different signatures.** The same data as in Figs. 3a,b of the main text, but in separate plots for the $EP_2$ locations determined by each of the four different signatures. **a,** Zeroes of the discriminant $D$. **b,** Phase vortices of the discriminant $D$. **c,** Zeroes of the eigenvector indicator $E$. **d,** Phase vortices of the eigenvector indicator $E$. The quantities $D$ and $E$ are defined in the main text, and additional discussion of $E$ is in the Supplementary Information. The projections used here are the same as in Figs. 3a,b of the main text. The solid curve is the same in all eight panels, and is the best-fit knot shown in Figs. 3a,b of the main text.

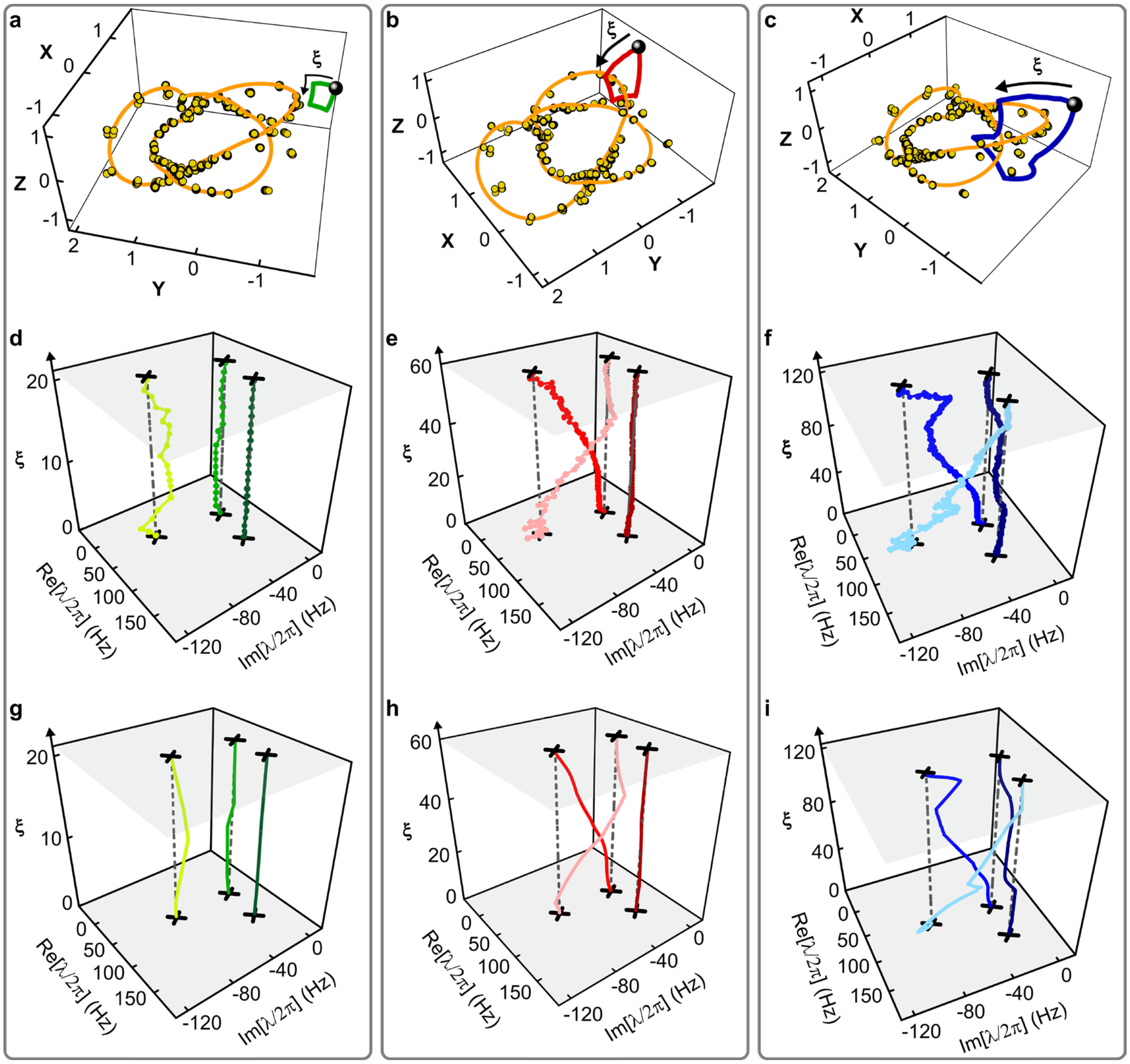

**Extended Data Fig. 6 | Comparison of measured and calculated braids. a-f,** The same panels as Fig. 3c-h of the main paper. They show the control loops (green, red, and blue in **a-c**) in relation to the measured knot (yellow circles) and the best-fit knot (orange curve). **d-f** show the resulting eigenvalue braids. **g-i,** The eigenvalue spectrum as calculated using the optomechanical parameters determined from fitting the knot of $EP_2$. The dashed lines are guides to the eye.

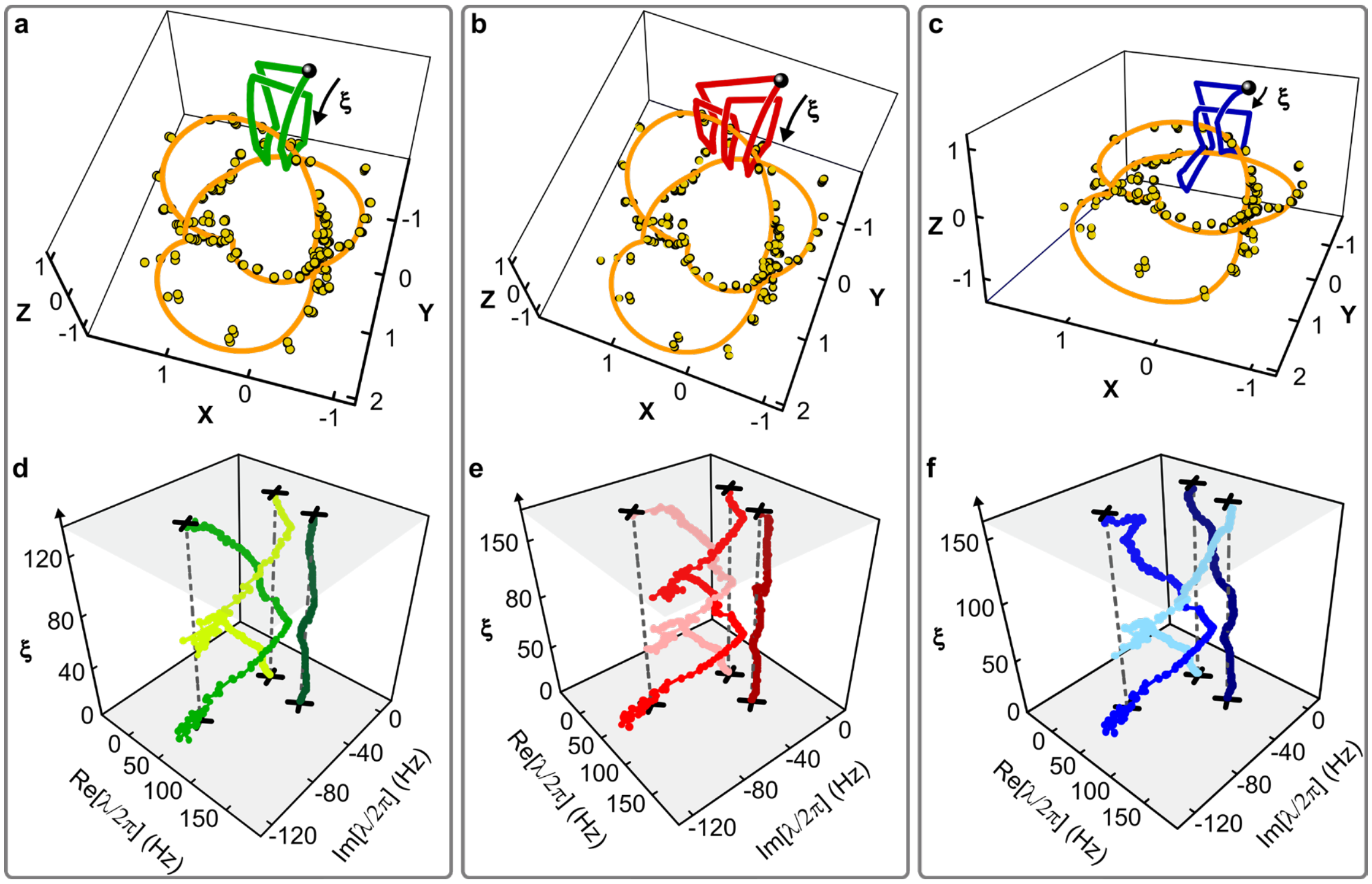

**Extended Data Fig. 7 | Additional braids of eigenvalues. a-c,** Three loops (green, red, blue), each from a different homotopy class. They share a common basepoint (black sphere) and are non-self-intersecting. The measured knot $\mathcal{K}$ (yellow circles) and the best-fit knot (orange curve) are shown for reference. The projection used here is the same as in Fig. 3a of the main text. **d-f,** the eigenvalue spectrum $\boldsymbol{\lambda}(\boldsymbol{\Psi})$ as $\boldsymbol{\Psi}$ is varied around a loop. The variable $\xi$ indexes the values of $\boldsymbol{\Psi}$ (along each loop) at which $\boldsymbol{\lambda}$ is measured. The black crosses show $\boldsymbol{\lambda}$ at the start and stop of the loop. The dashed lines are guides to the eye. The 1-$\sigma$ confidence intervals for $\boldsymbol{\lambda}$ are comparable to the size of the plotted points. The braids realized are: $\sigma_1^2$ (**d**), $\sigma_1^3$ (**e**), and $\sigma_2\sigma_1^2$ (**f**).

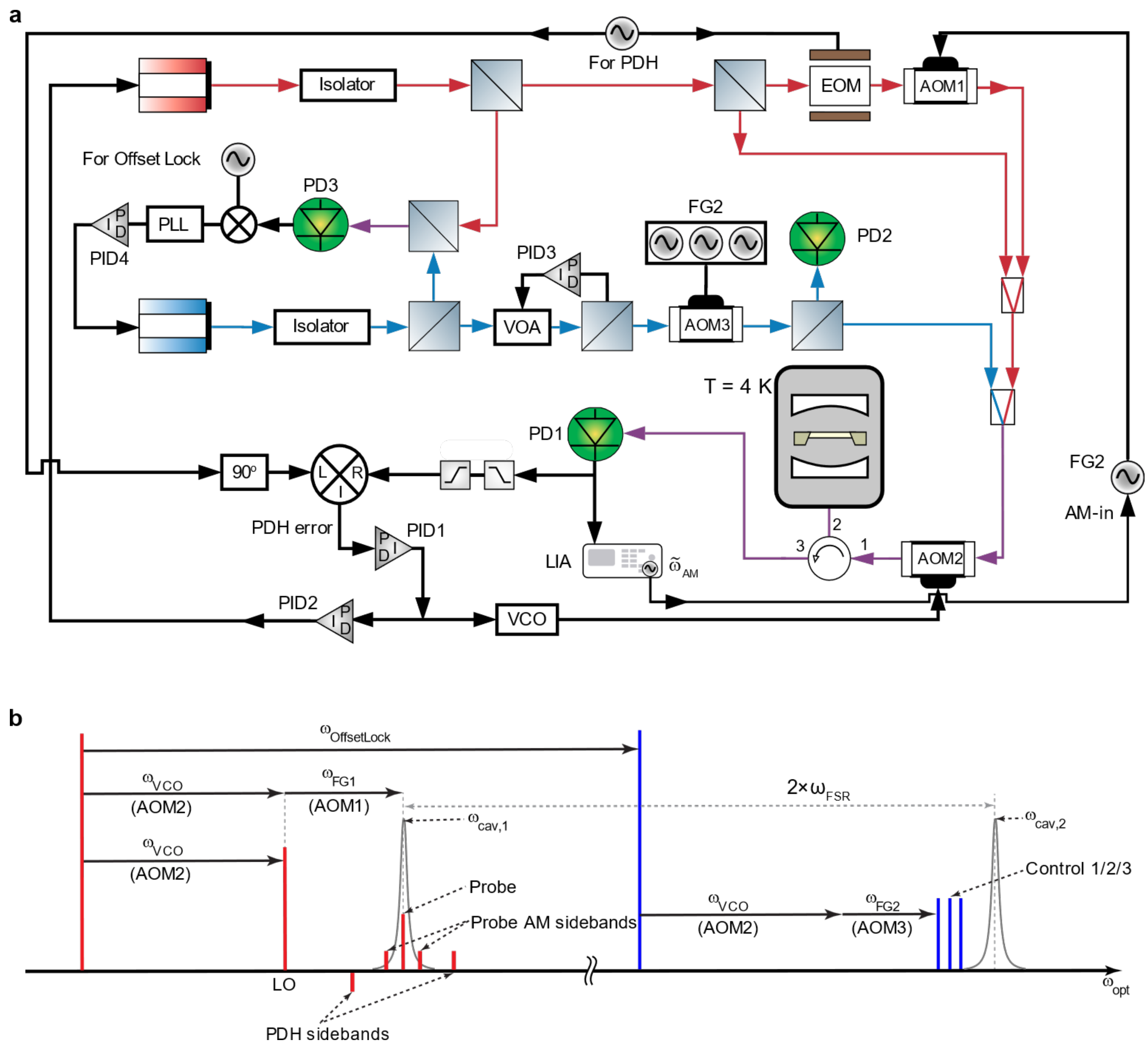

**Extended Data Fig. 8 | Details of the experimental setup. a,** The optical and electronic layout. Red arrows: beam path from the "probe" laser. Blue Arrows: beam path from the "control" laser. Purple arrows: overlapped beam path of the two lasers. Black arrows: electronic lines. Gray region: cryostat containing the optical cavity and membrane. The various components are described in the Supplementary Information. **b,** The optical spectrum. Red lines: tones produced from the probe laser. Blue lines: tones produced from the control laser. The tones and their generation are described in Methods and the Supplementary Information. Gray curves: the two cavity modes used in this work.

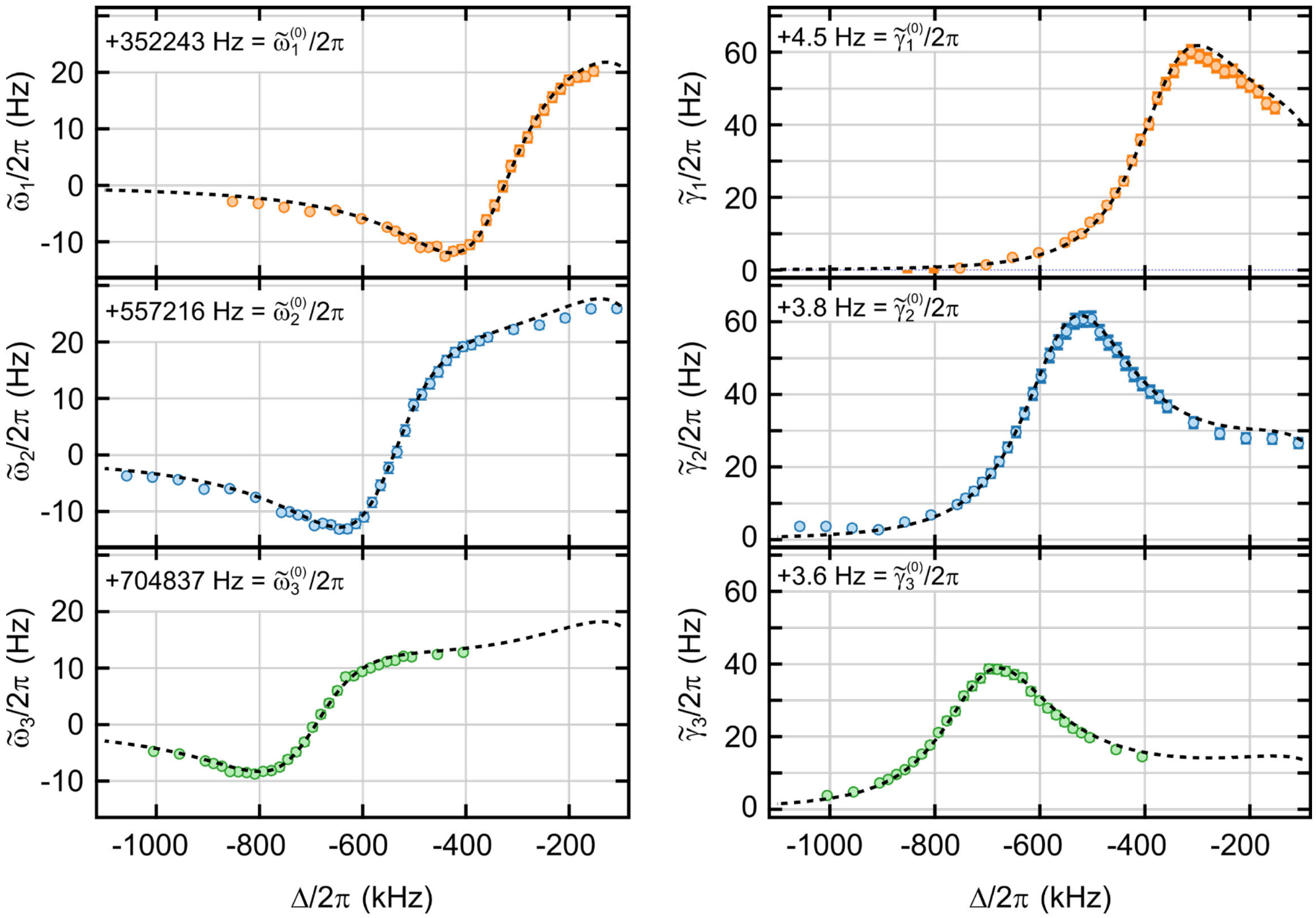

**Extended Data Fig. 9 | Characterizing the optomechanical coupling.** Here the cavity is driven with a single control tone, whose detuning (from the cavity resonance) is $\Delta$. Each panel shows the measured deviation of the (real or imaginary part of the) mechanical mode's eigenvalue from its bare value (i.e., from the relevant component of $\tilde{\boldsymbol{\lambda}}^{(0)}$, whose numerical value is written in the panel). The error bars show the 1-$\sigma$ confidence interval for each data point. A global fit to standard optomechanical theory gives the bare resonance frequencies $\tilde{\boldsymbol{\lambda}}^{(0)}$ and the optomechanical couplings $\boldsymbol{g}$. A detailed description of this procedure is in the Supplementary Information.

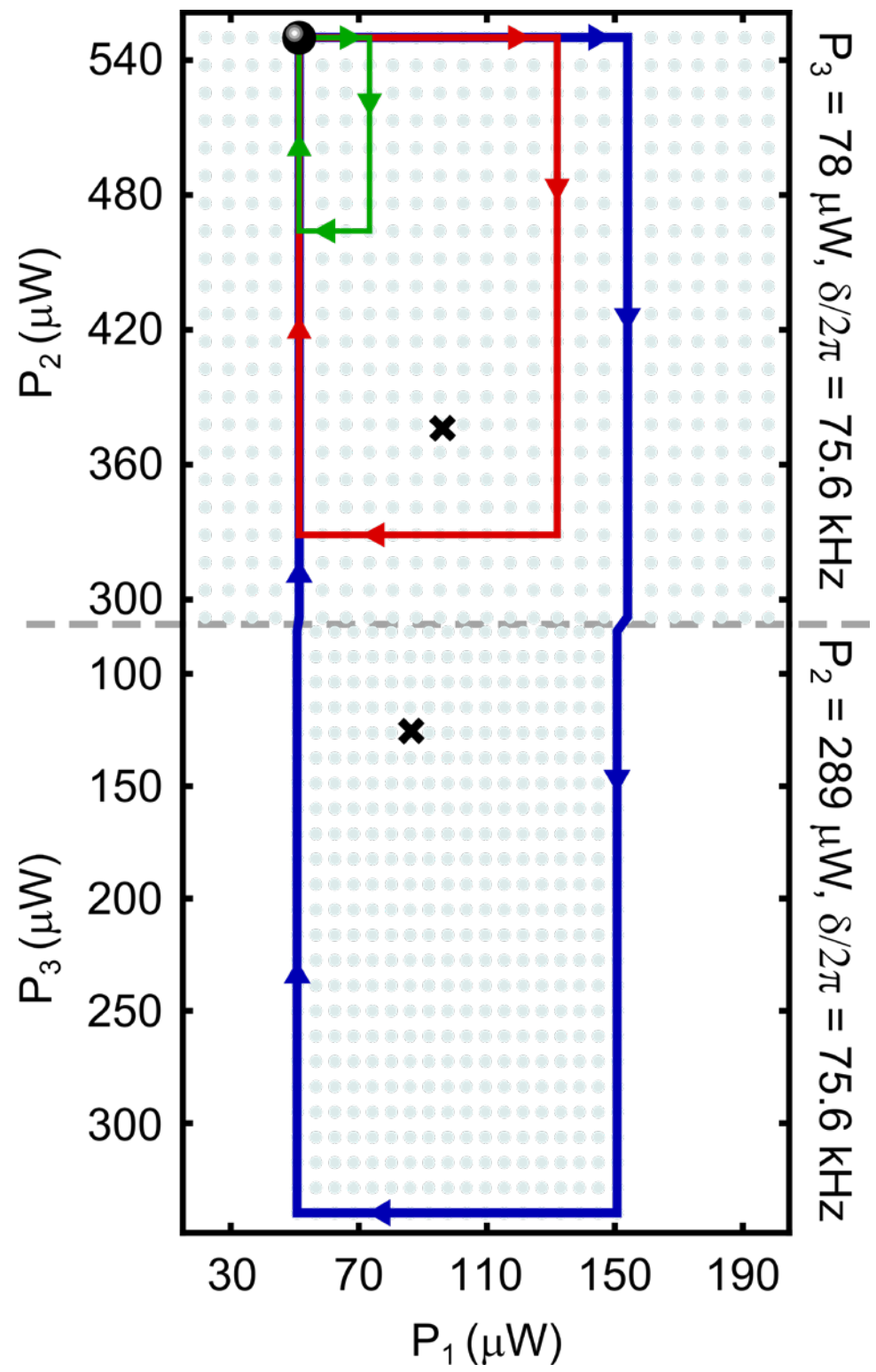

**Extended Data Fig. 10 | Control loops from Fig. 3c-e.** The three control loops in Fig. 3c-e of the main text were assembled from data taken in the two 2D sheets shown here. The two sheets' common border is shown as the dashed gray line. Each small gray disc represents a value of $\boldsymbol{\Psi}$ at which $\boldsymbol{\lambda}$ was measured (i.e., a "pixel" in the 2D sheet). The black crosses show the location of the $EP_2$ in these sheets as determined by the minima-finding algorithm described in the Supplementary Information.

# Supplementary Information for “Measuring the knot of non-Hermitian degeneracies and non-commuting braids”

## §0. Table of contents

## §1. Details of the experimental setup

This section describes the optical and electronic schemes used in this work.

The optical cavity is a single-sided Fabry-Perot resonator that is 3.64 cm long, for which the free spectral range (FSR) is measured to be $\omega_{FSR} = 2\pi \times 4.12$ GHz. It is built with mirrors having 5 cm radius of curvature and nominal reflectivities 0.9998 and 0.99997. With the membrane placed approximately at the center of the cavity, the finesse is $2.37 \times 10^4$. Fitting the cavity's reflection spectrum gives the input coupling efficiency $\kappa_{\text{in}}/\kappa = 0.267$.[i] The optomechanical cavity is mounted in a vacuum can which is placed inside a cryostat maintained at temperature $T = 4.2$ K, though the cryogenic temperature does not play a central role in this work. Details of the device's construction can be found in Refs. [ii,iii].

A detailed schematic of the extra-cavity setup is shown in Extended Data Fig. 8a. The optomechanical device is addressed by two Nd:YAG lasers (Innolight Prometheus), both operating with a wavelength of 1064 nm. The generation and use of the various optical tones derived from these lasers is described below. A schematic of the optical spectrum is depicted in Extended Data Fig. 8b.

### §1.1 *Laser 1 ("probe laser")*

Laser 1 is used to lock all of the optical tones relative to one of the cavity's resonances. It is also used to drive the mechanical modes and to detect their motion.

A portion of the light from Laser 1 is frequency-shifted by 279.5 MHz to generate a "probe" tone, using two acousto-optic modulators in series (AOM1 [Gooch & Housego free-space AOBD] and AOM2 [Gooch & Housego Fibre-Q]) driven at 79.5 MHz and 200 MHz, respectively. The probe tone is locked to an optical resonance of the cavity at $\omega_{\text{cav},1}$ using the Pound-Drever-Hall (PDH) technique[iv,v,vi]. This is accomplished using an electro-optic modulator (EOM [New Focus 4001]) driven at 15 MHz. A lock bandwidth of 4 kHz is realized by using the output from PID1 (Liquid Instruments Moku:Lab) to tune the voltage-controlled oscillator (VCO) that drives AOM2. Applying the feedback to AOM2 effectively locks all of the beams' detunings with respect to $\omega_{\text{cav},1}$, as discussed in §1.2.

To prevent large-amplitude, low-frequency drifts from forcing AOM2 beyond its effective tuning range, PID2 (Mokulabs) applies feedback control (with 1 Hz bandwidth) to the tuning piezo inside Laser 1 to maintain the VCO frequency close to 200 MHz.

Intensity modulation of the probe beam at frequencies $\widetilde{\omega}_{\text{AM}} \approx \widetilde{\omega}_{1,2,3}^{(0)}$ is produced using the amplitude-modulation (AM) input of the function generator FG1 (HP 4682B) that drives AOM1. It is this intensity modulation that drives the membrane (via radiation pressure) for the susceptibility measurements (an example of which is shown in Fig. 1c of the main paper and discussed in §4 of this supplement). The modulation is sourced from a lock-in amplifier (LIA [Zurich Instruments, HF2LI]) and modulates the probe beam intensity with a depth $\sim 0.04$.

Another portion of the light from Laser 1, which does not pass through AOM1, serves as the "local oscillator" (LO) for the heterodyne measurements.

The probe, its PDH and AM sidebands, and the LO are combined and sent to the optomechanical cavity's input port. These tones, together with the phase modulation sidebands induced on them by the membrane's motion, are then directed back from the optomechanical cavity through a circulator onto a photodetector (PD1 [Thorlabs PDA10CF, 150 MHz bandwidth]). The resulting photocurrent contains beatnotes at a variety of frequencies. To measure the beatnotes near 79.5 MHz, the output of PD1 is mixed down to 20.5 MHz (using a 100 MHz oscillator [Vaunix LSG121, not shown]), and then input to the LIA for phase-sensitive detection. For all of the measurements described here, the probe power is $\sim 15$ µW and the LO power is $\sim 1200$ µW, as measured at PD1.

**§1.2 *Laser 2 ("control laser")***

Laser 2 is used to generate the three "control" tones.

Laser 2 is locked to the unshifted light from Laser 1 with a frequency offset of $8234.098(3)$ MHz.[i] This frequency is chosen so that Laser 2 addresses a cavity mode whose resonance frequency $\omega_{\text{cav},2}$ differs from $\omega_{\text{cav},1}$ by $2\omega_{FSR}$ (i.e., whose longitudinal mode number differs by two from the mode addressed by Laser 1). The motivation for this approach is described below.

The light from Laser 2 is then frequency-shifted to generate the control tones 1,2,3. This shift is achieved using two AOMs in series. The first is AOM3 (Gooch & Housego, AOBD), which is driven by FG2 (2 units of Rigol DG4162) at three frequencies: 79.5 MHz – $\tilde{\omega}_1^{(0)} + \delta$, 79.5 MHz – $\tilde{\omega}_2^{(0)} + \eta + \delta$, and 79.5 MHz – $\tilde{\omega}_3^{(0)} + \delta$ (where $\delta$ is the "common detuning" that serves as one of the components of $\boldsymbol{\Psi}$). The second is AOM2 (which also controls the probe and LO), which is driven at 200 MHz.

As mentioned above, the PDH lock signal is applied to AOM2 to ensure that all of these beams track fluctuations of the cavity. As described in Refs. [ii,iii], the primary source of these fluctuations is low-frequency vibrations of the structure supporting the membrane chip. As a result, the two cavity modes will experience the same detuning only if they have the same optomechanical coupling to the membrane. To ensure that this is the case, the membrane's position within the cavity is chosen so that it lies at a "sweet spot", defined as a point where $\frac{d\omega_{\text{cav},1}}{dx_{\text{m}}} = \frac{d\omega_{\text{cav},2}}{dx_{\text{m}}}$, with $x_{\text{m}}$ being the membrane's position. The process for locating such a point is described in Refs. [vii,i,iii]

This also ensures that the optomechanical coupling between a given mechanical mode and any optical tone (whether produced by Laser 1 or Laser 2) is very nearly the same. In particular, the equality of the optomechanical coupling to the probe and control tones is useful in extracting the system's eigenvalues (see §2, §3, §4).

The powers of the control tones are stabilized (with a bandwidth of 1 kHz) using PID3 (New Focus LB1005), which feeds back to a variable optical attenuator (VOA, Thorlabs V1000A). The control tone powers reported in this work are all measured at PD2 (Thorlabs PDA36A), and range from 0 µW to 800 µW.

**§2. Optomechanical characterization**

This section details the measurement of the optomechanical coupling rates $\boldsymbol{g}$ and the bare mechanical eigenvalues $\tilde{\boldsymbol{\lambda}}^{(0)}$. Briefly, these are obtained by measuring the dynamical back-action (DBA) for each mechanical mode when the cavity is driven by a single control tone (i.e., rather than the three tones used in the main part of this work). Note that frequencies associated with the mechanical modes are written with a tilde (e.g. $\tilde{\boldsymbol{\lambda}}$) when they are given in the lab frame, and without a tilde (e.g., $\boldsymbol{\lambda}$) when they are given in the rotating frame $\mathcal{R}$ (which is defined in §3).

To characterize mode $i$ of the membrane (where the modes are indexed by $i \in \{1,2,3\}$), the optical cavity is driven by a single control tone with power $\sim 250$ µW and detuning $\Delta$, which is stepped over a range $\sim 3\kappa$ centered near $-\tilde{\omega}_i^{(0)}$. At each value of $\Delta$, the mechanical susceptibility of the mode is measured (as described in §4) and fit to a complex Lorentzian to yield the resonance frequency $\tilde{\omega}_i \equiv \mathrm{Re}(\tilde{\lambda}_i)$ and energy damping rate $\tilde{\gamma}_i \equiv -2\mathrm{Im}(\tilde{\lambda}_i)$, both of which are tuned via dynamical back-action (DBA). This is illustrated in Extended Data Fig. 9.

For $\Delta \ll -\tilde{\omega}_i^{(0)}$, the DBA is expected to approach zero. Thus, the asymptotes in Extended Data Fig. 9 correspond to the bare mechanical eigenvalues $\tilde{\boldsymbol{\lambda}}^{(0)}$. In addition, the DBA-induced shift in $\tilde{\omega}_i$ is expected to be zero (and the shift in $\tilde{\gamma}_i$ is expected to be a maximum) at $\Delta \approx -\tilde{\omega}_i^{(0)}$. This feature is helpful in identifying any spurious shift $\Delta_0$ in the laser detuning. The value of $\Delta_0$ found from fitting data as in Extended Data Fig. 9 is typically less than $2\pi \times 5$ kHz. Once $\Delta_0$ is measured, it is compensated by adding a corresponding shift to the offset lock of Laser 2 (which generates the three control tones, see §1).

The best-fit values of $\boldsymbol{g}$ and $\tilde{\boldsymbol{\lambda}}^{(0)}$ are determined from a global fit of the dataset shown in Extended Data Fig. 9 to standard optomechanics theory.[34] These values are:

$$\tilde{\boldsymbol{\lambda}}^{(0)} = 2\pi \times (352243.3^{\pm 0.1} - 2.2^{\pm 0.1} i, 557216.8^{\pm 0.1} - 1.9^{\pm 0.1} i, 704836.7^{\pm 0.1} - 1.8^{\pm 0.1} i) \text{ Hz}$$
$$\boldsymbol{g} = 2\pi \times (0.198^{\pm 0.001}, 0.304^{\pm 0.001}, 0.300^{\pm 0.001}) \text{ Hz}$$

where the uncertainties indicate one standard deviation.

Note that we define $\boldsymbol{g} = \sqrt{\eta_c}\, \boldsymbol{g_0}$, where $\boldsymbol{g_0}$ is the conventionally reported single-photon optomechanical coupling rate. Here the coupling efficiency $\eta_c = P_{\mathrm{in}}/P_{\mathrm{meas}}$, where $P_{\mathrm{in}}$ is the optical power incident on the optomechanical device and $P_{\mathrm{meas}}$ is the power detected at PD2. Since the absolute magnitude of $\boldsymbol{g}_0$ is inconsequential to the results presented in this work, no attempt is made to calibrate $\eta_c$. However, for completeness we note that the theoretically expected value of $\boldsymbol{g_0} = 2\pi \times (5.5, 4.3, 3.9)$ Hz.

## §3. Modelling the three-mode system

This section describes the theoretical model for optically controlling **H,** the dynamical matrix of the mechanical three-mode system. It also defines the rotating frame $\mathcal{R}$ in which the bare mechanical modes are nearly degenerate, and in which the EPs are accessed. Lastly, it details how the experimental parameters that correspond to an $EP_3$ are determined, and how it is verified that these parameters span the space of spectra in the neighborhood of this $EP_3$.

### §3.1 *Optomechanical model*

The classical Hamiltonian function for the full optomechanical system (i.e., including the optical mode as well as the three mechanical modes) is:

$$\boldsymbol{\mathcal{H}} = \hbar\big(\Omega_{\text{cav},2} - i\,\kappa/2\big)a^* a + \sum_{i=1}^{3} \hbar\tilde{\lambda}_i^{(0)} \tilde{c}_i^* \tilde{c}_i - \sum_{i=1}^{3} \hbar g_i (\tilde{c}_i^* + \tilde{c}_i) a^* a$$

where $a$ is the complex amplitude of the optical mode addressed by the control laser, the $\tilde{c}_i$ are the complex amplitudes of the three mechanical modes, and * denotes complex conjugation. The first two terms correspond to the uncoupled optical and mechanical oscillators, and the third term corresponds to their interaction (with coupling strengths $g_i$). Note that the expression for $\boldsymbol{\mathcal{H}}$ includes the reduced Planck's constant $\hbar$ only in order to conform with the broader literature on optomechanics, in which the coupling rates would be the single-photon rates and the mode amplitudes would be the corresponding quantum mechanical operators. Since this work is purely classical, the overall scale of $\boldsymbol{\mathcal{H}}$ (and hence the appearance of $\hbar$) is irrelevant.

The dynamics of $a$ and $\tilde{c}_i$ are governed by $\boldsymbol{\mathcal{H}}$ via Hamilton's equations. The optical cavity is also driven through its input port (with coupling $\kappa_{\text{in}}$) with a drive field

$$a_{\text{in}}(t) = \sum_{j=1}^{3} \sqrt{\frac{P_j}{\hbar\Omega_j}}\, e^{-i(\Omega_j t + \phi_j)}$$

corresponding to the three control tones with powers $P_j$, frequencies $\Omega_j$ (i.e. detunings $\Delta_j = \Omega_j - \omega_{\text{cav},2}$), and phases $\phi_j$. Because $\kappa \gg g_i$, the interaction term in $\boldsymbol{\mathcal{H}}$ can be linearized with respect to $a$. And since $\kappa \gg \tilde{\gamma}_i$, the optical field can be adiabatically eliminated to yield an effective equation of motion for the mechanical modes, in which $a$ does not appear but in which the parameters of the cavity drive (i.e., of the control tones) do.

For the detunings used in this work (see Fig. 2b of the main text):

$$\Delta_1 = -\widetilde{\omega}_1^{(0)} + \delta, \qquad \Delta_2 = -\widetilde{\omega}_2^{(0)} + \delta + \eta, \qquad \Delta_3 = -\widetilde{\omega}_3^{(0)} + \delta$$

(with $\eta = -2\pi \times 100$ Hz), the intensity beatnote between control tones $j$ and $k$ is at the frequency

$$\left|\Delta_{jk}\right| = \left|\Delta_j - \Delta_k\right| \approx \left|\widetilde{\omega}_j^{(0)} - \widetilde{\omega}_k^{(0)}\right|.$$

Under the rotating wave approximation, the control tones $j$ and $k$ thus couple only the two mechanical modes $j$ and $k$, and not the third mechanical mode. As a result, the equation of motion for the three mechanical modes reduces to the form

$$\dot{\tilde{\boldsymbol{c}}} = \begin{pmatrix} \dot{\tilde{c}}_1 \\ \dot{\tilde{c}}_2 \\ \dot{\tilde{c}}_3 \end{pmatrix} = -i\,\widetilde{\mathbf{H}} \begin{pmatrix} \tilde{c}_1 \\ \tilde{c}_2 \\ \tilde{c}_3 \end{pmatrix} = -i\widetilde{\mathbf{H}}\,\tilde{\boldsymbol{c}}$$

where

$$\widetilde{\mathbf{H}} = \begin{pmatrix} \tilde{\lambda}_1^{(0)} & 0 & 0 \\ 0 & \tilde{\lambda}_2^{(0)} & 0 \\ 0 & 0 & \tilde{\lambda}_3^{(0)} \end{pmatrix} + \begin{pmatrix} \sigma_{11} & \sigma_{12} e^{i\Delta_{12} t} e^{i\phi_{12}} & \sigma_{13} e^{i\Delta_{13} t} e^{i\phi_{13}} \\ \sigma_{21} e^{-i\Delta_{12} t} e^{-i\phi_{12}} & \sigma_{22} & \sigma_{23} e^{i\Delta_{23} t} e^{i\phi_{23}} \\ \sigma_{31} e^{-i\Delta_{13} t} e^{-i\phi_{13}} & \sigma_{32} e^{-i\Delta_{23} t} e^{-i\phi_{23}} & \sigma_{33} \end{pmatrix}$$

and $\phi_{jk} = \phi_j - \phi_k$.

The coefficients denoted by $\sigma$ are time-independent, and depend on the parameters of the optical drive. Specifically, the off-diagonal components are given by:

$$\sigma_{jk} = -i\kappa_{\text{in}}\, g_j g_k \left[\sqrt{\frac{P_j}{\hbar\Omega_j}\frac{P_k}{\hbar\Omega_k}}\,\chi_{\text{cav}}^*(\Delta_j)\chi_{\text{cav}}(\Delta_k)\left[\chi_{\text{cav}}\left(\widetilde{\omega}_j^{(0)} + \Delta_j\right) - \chi_{\text{cav}}\left(\widetilde{\omega}_j^{(0)} - \Delta_k\right)\right]\right]$$

and the diagonal components are given by:

$$\sigma_{jj} = -i\kappa_{\text{in}}\, g_j^2\, \sum_{k=1,2,3}\left[\frac{P_k}{\hbar\Omega_k}\left|\chi_{\text{cav}}(\Delta_k)\right|^2\left[\chi_{\text{cav}}\left(\widetilde{\omega}_j^{(0)} + \Delta_k\right) - \chi_{\text{cav}}\left(\widetilde{\omega}_j^{(0)} - \Delta_k\right)\right]\right]$$

where the cavity's optical susceptibility is

$$\chi_{\text{cav}}(\Delta) = \frac{1}{\kappa/2 - i\Delta}$$

We can remove the explicit time dependence from $\widetilde{\mathbf{H}}$ by writing the equation of motion in the rotating frame $\mathcal{R}$ defined by the transformation $\mathbf{U}$ given immediately below. In this frame the equation of motion is

$$\dot{\boldsymbol{c}} = -i\,\mathbf{H}\,\boldsymbol{c}$$

where

$$\boldsymbol{c} = \begin{pmatrix} c_1(t) \\ c_2(t) \\ c_3(t) \end{pmatrix} = \mathbf{U} \begin{pmatrix} \tilde{c}_1(t) \\ \tilde{c}_2(t) \\ \tilde{c}_3(t) \end{pmatrix} = \begin{pmatrix} e^{i\left(\tilde{\omega}_1^{(0)}+\eta\right)t} e^{-i\phi_1} & 0 & 0 \\ 0 & e^{i\tilde{\omega}_2^{(0)}t} e^{-i\phi_2} & 0 \\ 0 & 0 & e^{i\left(\tilde{\omega}_3^{(0)}+\eta\right)t} e^{-i\phi_3} \end{pmatrix} \begin{pmatrix} \tilde{c}_1(t) \\ \tilde{c}_2(t) \\ \tilde{c}_3(t) \end{pmatrix}$$

and the dynamical matrix $\mathbf{H} = \mathbf{U}\widetilde{\mathbf{H}}\mathbf{U}^{-1} + i\dot{\mathbf{U}}\mathbf{U}^{-1}$ is time-independent:

$$\mathbf{H} = \begin{pmatrix} -\eta - i\tilde{\gamma}_1^{(0)}/2 & 0 & 0 \\ 0 & -i\tilde{\gamma}_2^{(0)}/2 & 0 \\ 0 & 0 & -\eta - i\tilde{\gamma}_3^{(0)}/2 \end{pmatrix} + \begin{pmatrix} \sigma_{11} & \sigma_{12} & \sigma_{13} \\ \sigma_{21} & \sigma_{22} & \sigma_{23} \\ \sigma_{31} & \sigma_{32} & \sigma_{33} \end{pmatrix} \qquad \text{(S1)}$$

It is in this frame $\mathcal{R}$ that $\mathbf{H}$ can be brought to an $EP_3$ degeneracy by controlling the second matrix in Eq. S1 (denoted $\boldsymbol{\sigma}$) with $\boldsymbol{\Psi} = (\delta, P_1, P_2, P_3)$.

### §3.2 *Locating an $EP_3$*

The dynamical matrix described by Eq. S1 is a complicated function of the experimental control parameters $(\delta, P_1, P_2, P_3)$ and we did not find an analytic means for determining values of these parameters that correspond to a three-fold degeneracy. Instead, we searched numerically over a wide range of these parameters. This search revealed several three-fold degeneracies. The one used for this work (i.e., $\boldsymbol{\Psi}_{\mathrm{EP3}}$) was chosen because it corresponded to the most readily accessible values of $(\delta, P_1, P_2, P_3)$.

### §3.3 *Spanning the neighborhood of $EP_3$*

To test whether the four experimental parameters $(\delta, P_1, P_2, P_3)$ span the space of spectra around an $EP_3$ degeneracy, we use the inverse function theorem to argue about the existence of a map between these four parameters and the two complex coefficients $(x, y)$ of the characteristic polynomial of $\mathbf{H_0}$ in the vicinity of $\boldsymbol{\Psi}_{\mathrm{EP3}}$. For simplicity, here we use $\mathbf{H_0}$ (the traceless version of $\mathbf{H}$, defined as $\mathbf{H_0} = \mathbf{H} - \mathrm{tr}(\mathbf{H})\mathbf{I}/3$ where $\mathbf{I}$ is the identity matrix), in which case $x = \det(\mathbf{H}_0)$ and $y = \mathrm{tr}(\mathbf{H}_0^2)/2$.

In particular, we consider the Jacobian $\mathbf{J}$ of this map, where

$$\mathbf{J} = \begin{pmatrix} \partial\mathrm{Re}(x)/\partial\delta & \partial\mathrm{Re}(x)/\partial P_1 & \partial\mathrm{Re}(x)/\partial P_2 & \partial\mathrm{Re}(x)/\partial P_3 \\ \partial\mathrm{Im}(x)/\partial\delta & \partial\mathrm{Im}(x)/\partial P_1 & \partial\mathrm{Im}(x)/\partial P_2 & \partial\mathrm{Im}(x)/\partial P_3 \\ \partial\mathrm{Re}(y)/\partial\delta & \partial\mathrm{Re}(y)/\partial P_1 & \partial\mathrm{Re}(y)/\partial P_2 & \partial\mathrm{Re}(y)/\partial P_3 \\ \partial\mathrm{Im}(y)/\partial\delta & \partial\mathrm{Im}(y)/\partial P_1 & \partial\mathrm{Im}(y)/\partial P_2 & \partial\mathrm{Im}(y)/\partial P_3 \end{pmatrix}$$

and the derivatives are evaluated at $\boldsymbol{\Psi}_{\mathrm{EP3}}$. $\mathbf{J}$ is continuously differentiable in $\boldsymbol{\Psi}$ because $x$ and $y$ are smooth functions in the elements of $\mathbf{H_0}$, which in turn are continuously differentiable in $\boldsymbol{\Psi}$ (over the range of $\boldsymbol{\Psi}$ used in these measurements). Therefore, if $\det(\mathbf{J}) \neq 0$, the parameters span the same space as $x$ and $y$ (which is the full space of spectra, as discussed in the main paper) in the neighborhood of the $\mathrm{EP_3}$.

Numerical evaluation of $\det(\mathbf{J})$ is carried out using the expression for $\mathbf{H}$ in Eq. S1 (and the location of $\boldsymbol{\Psi}_{\mathrm{EP3}}$ as determined in Methods), giving $\det(\mathbf{J}) \approx 10^{30}\ (2\pi\ \mathrm{Hz})^9/\mathrm{W}^3$. This value is non-zero. More precisely, it is of the order of magnitude expected from the form of $\mathbf{J}$. It is roughly equal to $\left(\lambda^{(\mathrm{typ})}\right)^{10}/(\Delta P)^3(\Delta\delta)$ where $\lambda^{(\mathrm{typ})}$ is the typical magnitude of the eigenvalues in the neighborhood of $\boldsymbol{\Psi}_{\mathrm{EP3}}$, and $\Delta P$ and $\Delta\delta$ are the typical scales of the control parameters over which the $\lambda_i$ vary.

The connection between the experimental control parameters and the coefficients of the characteristic polynomial (in the neighborhood of $\mathrm{EP_3}$) can also be understood by noting that any traceless matrix in the neighborhood of an $\mathrm{EP_3}$ can be brought (via similarity transformation) to the canonical form (known as Arnol'd-Jordan normal form)[6,7]

$$\underline{\mathbf{H_0}} = \begin{pmatrix} 0 & 1 & 0 \\ 0 & 0 & 1 \\ x & y & 0 \end{pmatrix}$$

where $x$ and $y$ are the two complex coefficients of the characteristic polynomial, as described in the preceding paragraphs (and in §1). $\underline{\mathbf{H_0}}$ may also be regarded as the companion matrix of the characteristic polynomial. This form highlights the fact that the non-vanishing of $\det(\mathbf{J})$ (described in the preceding paragraphs) ensures the existence of a linear relationship between the experimental control parameters and the matrix elements of $\mathbf{H_0}$.

**§4. Extracting the spectrum from mechanical susceptibility measurements**

This section describes the relationship between the system's eigenvalue spectrum $\boldsymbol{\lambda}$ and measurements of the mechanical susceptibility. In particular, it derives the functional form used to fit the susceptibility data (e.g., as shown in Fig. 1c of the main text).

In the rotating frame $\mathcal{R}$ (see §3), the mechanical modes' response to a force $\boldsymbol{f}(\omega)$ can be written in the Fourier domain as

$$\boldsymbol{c}(\omega) = \boldsymbol{\chi}(\omega)\boldsymbol{f}(\omega)$$

where

$$\boldsymbol{\chi}(\omega) = (\omega\mathbf{I} - \mathbf{H})^{-1}$$

The principle behind the measurements used in this work is to apply a force $\boldsymbol{f}(\omega)$, measure the mechanical response $\boldsymbol{c}(\omega)$, and thus infer the susceptibility $\boldsymbol{\chi}(\omega)$, which contains information about $\boldsymbol{\lambda}$.

The measurement of the mechanical response is carried out in the lab frame, where

$$\tilde{\boldsymbol{c}}(\widetilde{\omega}) = \begin{pmatrix} \tilde{c}_1(\widetilde{\omega}) \\ \tilde{c}_2(\widetilde{\omega}) \\ \tilde{c}_3(\widetilde{\omega}) \end{pmatrix} = \begin{pmatrix} c_1\left(\widetilde{\omega} - \widetilde{\omega}_1^{(0)} - \eta\right) \\ c_2\left(\widetilde{\omega} - \widetilde{\omega}_2^{(0)}\right) \\ c_3\left(\widetilde{\omega} - \widetilde{\omega}_3^{(0)} - \eta\right) \end{pmatrix}$$

$$= \begin{pmatrix} \sum_{j=1}^{3} \left[\boldsymbol{\chi}\left(\widetilde{\omega} - \widetilde{\omega}_1^{(0)} - \eta\right)\right]_{1,j} \left[\boldsymbol{f}\left(\widetilde{\omega} - \widetilde{\omega}_1^{(0)} - \eta\right)\right]_j \\ \sum_{j=1}^{3} \left[\boldsymbol{\chi}\left(\widetilde{\omega} - \widetilde{\omega}_2^{(0)}\right)\right]_{2,j} \left[\boldsymbol{f}\left(\widetilde{\omega} - \widetilde{\omega}_2^{(0)}\right)\right]_j \\ \sum_{j=1}^{3} \left[\boldsymbol{\chi}\left(\widetilde{\omega} - \widetilde{\omega}_3^{(0)} - \eta\right)\right]_{3,j} \left[\boldsymbol{f}\left(\widetilde{\omega} - \widetilde{\omega}_3^{(0)} - \eta\right)\right]_j \end{pmatrix}$$

The force applied by the intensity modulation of the probe tone is (in the lab frame) $\tilde{\boldsymbol{f}}(t) \propto e^{i\widetilde{\omega}_{\mathrm{AM}}t}\boldsymbol{g}$, i.e. in the Fourier domain $\tilde{\boldsymbol{f}}(\widetilde{\omega}) \propto \boldsymbol{g}\,\delta(\widetilde{\omega} - \widetilde{\omega}_{\mathrm{AM}})$. In $\mathcal{R}$ this is:

$$\boldsymbol{f}(\omega) \propto \begin{pmatrix} g_1\delta\left(\omega - \widetilde{\omega}_{\mathrm{AM}} + \widetilde{\omega}_1^{(0)} + \eta\right) \\ g_2\delta\left(\omega - \widetilde{\omega}_{\mathrm{AM}} + \widetilde{\omega}_2^{(0)}\right) \\ g_3\delta\left(\omega - \widetilde{\omega}_{\mathrm{AM}} + \widetilde{\omega}_3^{(0)} + \eta\right) \end{pmatrix}$$

Thus, driving the membrane with a single sinusoidal force results in motion at three different frequencies. However, the lock-in amplifier only detects motion at the drive frequency $\tilde{\omega}_{\mathrm{AM}}$, i.e.

$$\tilde{V}[\tilde{\omega}_{\mathrm{AM}}] = \alpha \int f_{\mathrm{LIA}}(\tilde{\omega} - \tilde{\omega}_{\mathrm{AM}}) \tilde{\boldsymbol{c}}(\tilde{\omega}) \cdot \boldsymbol{g}\, d\tilde{\omega} \approx \alpha \int_{\tilde{\omega}_{\mathrm{AM}}-\xi}^{\tilde{\omega}_{\mathrm{AM}}+\xi} \tilde{\boldsymbol{c}}(\tilde{\omega}) \cdot \boldsymbol{g}\, d\tilde{\omega}$$

where $f_{\mathrm{LIA}}(x)$ is the filter function of the lock-in amplifier (which has effective bandwidth $\xi$), and $\alpha$ is the transduction gain. As a result,

$$\tilde{V}(\tilde{\omega}_{\mathrm{AM}}) = \begin{cases} \alpha\, g_1^2 \left[\boldsymbol{\chi}\left(\tilde{\omega}_{\mathrm{AM}} - \tilde{\omega}_1^{(0)} - \eta\right)\right]_{1,1} & \equiv \tilde{V}_1(\tilde{\omega}_{\mathrm{AM}}) \qquad \text{for } \tilde{\omega}_{\mathrm{AM}} \approx \tilde{\omega}_1^{(0)} \\ \alpha\, g_2^2 \left[\boldsymbol{\chi}\left(\tilde{\omega}_{\mathrm{AM}} - \tilde{\omega}_2^{(0)}\right)\right]_{2,2} & \equiv \tilde{V}_2(\tilde{\omega}_{\mathrm{AM}}) \qquad \text{for } \tilde{\omega}_{\mathrm{AM}} \approx \tilde{\omega}_2^{(0)} \\ \alpha\, g_3^2 \left[\boldsymbol{\chi}\left(\tilde{\omega}_{\mathrm{AM}} - \tilde{\omega}_3^{(0)} - \eta\right)\right]_{3,3} & \equiv \tilde{V}_3(\tilde{\omega}_{\mathrm{AM}}) \qquad \text{for } \tilde{\omega}_{\mathrm{AM}} \approx \tilde{\omega}_3^{(0)} \end{cases}$$

so that only the diagonal components of the susceptibility $\boldsymbol{\chi}(\omega)$ are measured. Each of these diagonal components contains $\boldsymbol{\lambda}$, so in principle it would suffice to measure $\tilde{V}(\tilde{\omega}_{\mathrm{AM}})$ in just one of the frequency ranges (say, around $\tilde{\omega}_2^{(0)}$). However, to make the analysis robust against noise, $\tilde{V}(\tilde{\omega}_{\mathrm{AM}})$ was measured in all three frequency ranges (i.e., around each of the $\tilde{\omega}_{1,2,3}^{(0)}$), and this nominally redundant data was fit to determine $\boldsymbol{\lambda}$.

To explicitly see the relation of the susceptibility $\boldsymbol{\chi}(\omega)$ to the eigenspectrum of $\mathbf{H}$, consider its diagonalization $\mathbf{H} = \mathbf{TDT^{-1}}$, where

$$\mathbf{D} = \begin{pmatrix} \lambda_1 & 0 & 0 \\ 0 & \lambda_2 & 0 \\ 0 & 0 & \lambda_3 \end{pmatrix}.$$

It can be easily shown that $\boldsymbol{\chi}(\omega) = \mathbf{T}(\omega\mathbf{I} - \mathbf{D})^{-1}\mathbf{T}^{-1}$, where $(\omega\mathbf{I} - \mathbf{D})^{-1}$ is diagonal and contains $\boldsymbol{\lambda}$ as

$$(\omega\mathbf{I} - \mathbf{D})^{-1} = \begin{pmatrix} \dfrac{1}{\omega - \lambda_1} & 0 & 0 \\ 0 & \dfrac{1}{\omega - \lambda_2} & 0 \\ 0 & 0 & \dfrac{1}{\omega - \lambda_3} \end{pmatrix}.$$

This can be used to write the $\tilde{V}_i(\tilde{\omega}_{\mathrm{AM}})$ explicitly in terms of $\boldsymbol{\lambda}$ and the matrix elements of $\mathbf{T}$ and $\mathbf{T}^{-1}$ as

$$\tilde{V}_1(\tilde{\omega}) = \alpha\, g_1^2 \left[\frac{T_{11}(T^{-1})_{11}}{\tilde{\omega} - \tilde{\omega}_1^{(0)} - \eta - \lambda_1} + \frac{T_{12}(T^{-1})_{21}}{\tilde{\omega} - \tilde{\omega}_1^{(0)} - \eta - \lambda_2} + \frac{T_{13}(T^{-1})_{31}}{\tilde{\omega} - \tilde{\omega}_1^{(0)} - \eta - \lambda_3}\right]$$
$$\tilde{V}_2(\tilde{\omega}) = \alpha\, g_2^2 \left[\frac{T_{21}(T^{-1})_{12}}{\tilde{\omega} - \tilde{\omega}_2^{(0)} - \lambda_1} + \frac{T_{22}(T^{-1})_{22}}{\tilde{\omega} - \tilde{\omega}_2^{(0)} - \lambda_2} + \frac{T_{23}(T^{-1})_{32}}{\tilde{\omega} - \tilde{\omega}_2^{(0)} - \lambda_3}\right]$$
$$\tilde{V}_3(\tilde{\omega}) = \alpha\, g_3^2 \left[\frac{T_{31}(T^{-1})_{13}}{\tilde{\omega} - \tilde{\omega}_3^{(0)} - \eta - \lambda_1} + \frac{T_{32}(T^{-1})_{23}}{\tilde{\omega} - \tilde{\omega}_3^{(0)} - \eta - \lambda_2} + \frac{T_{33}(T^{-1})_{33}}{\tilde{\omega} - \tilde{\omega}_3^{(0)} - \eta - \lambda_3}\right]$$

To extract the eigenvalues $\boldsymbol{\lambda}$, these three spectra can be fit to the sum of nine complex Lorentzians as

$$\tilde{V}_1(\tilde{\omega}) = a_1 \left[\frac{s_{11}}{\tilde{\omega} - \tilde{\omega}_1^{(0)} - \eta - \lambda_1} + \frac{s_{12}}{\tilde{\omega} - \tilde{\omega}_1^{(0)} - \eta - \lambda_2} + \frac{s_{13}}{\tilde{\omega} - \tilde{\omega}_1^{(0)} - \eta - \lambda_3}\right] + b_1$$
$$\tilde{V}_2(\tilde{\omega}) = a_2 \left[\frac{s_{21}}{\tilde{\omega} - \tilde{\omega}_2^{(0)} - \lambda_1} + \frac{s_{22}}{\tilde{\omega} - \tilde{\omega}_2^{(0)} - \lambda_2} + \frac{s_{23}}{\tilde{\omega} - \tilde{\omega}_2^{(0)} - \lambda_3}\right] + b_2$$
$$\tilde{V}_3(\tilde{\omega}) = a_3 \left[\frac{s_{31}}{\tilde{\omega} - \tilde{\omega}_3^{(0)} - \eta - \lambda_1} + \frac{s_{32}}{\tilde{\omega} - \tilde{\omega}_3^{(0)} - \eta - \lambda_2} + \frac{s_{33}}{\tilde{\omega} - \tilde{\omega}_3^{(0)} - \eta - \lambda_3}\right] + b_3$$

where $a_i = \alpha\, g_i^2$ and $s_{ij} = T_{ij}(T^{-1})_{ji}$, and the three additional (complex) constants $b_i$ represent the lock-in-detection background. Of the 18 complex parameters in this model ($a_i, b_i, s_{ij}$, and $\lambda_i$), the amplitudes $s_{ij}$ are constrained by the fact that $\mathbf{T}\mathbf{T}^{-1} = \mathbf{I} = \mathbf{T}^{-1}\mathbf{T}$, i.e. $\sum_j T_{ij}(T^{-1})_{ji} = 1 = \sum_j (T^{-1})_{ij}T_{ji}$, which implies that $\sum_j s_{ij} = 1 = \sum_j s_{ji}$. Therefore, the rows and columns of the matrix

$$\mathbf{S} = \begin{pmatrix} s_{11} & s_{12} & s_{13} \\ s_{21} & s_{22} & s_{23} \\ s_{31} & s_{32} & s_{33} \end{pmatrix}$$

each add to unity. These are five independent complex constraints, and are implemented in fitting the measured spectra as:

$$s_{13} = 1 - s_{11} - s_{12}$$
$$s_{23} = 1 - s_{21} - s_{22}$$
$$s_{31} = 1 - s_{11} - s_{21}$$
$$s_{32} = 1 - s_{12} - s_{22}$$
$$s_{33} = s_{11} + s_{12} + s_{21} + s_{22} - 1.$$

In other words, the global fit of the measured spectra $\tilde{V}_1[\widetilde{\omega}]$, $\tilde{V}_2[\widetilde{\omega}], \tilde{V}_3[\widetilde{\omega}]$ to nine complex Lorentzians is implemented with 13 complex fit parameters. The best-fit values for $\boldsymbol{\lambda}$ and **S** so extracted are used in various ways to identify the locations of $EP_2$ and $EP_3$ (see §7).

**§5. Degeneracy of eigenvectors**

This section describes the connection between eigenvector degeneracy and the vanishing of $E = (\det(\mathbf{S}))^{-2}$.

We first note that in the susceptibility measurements, the membrane motion is detected only at the actuation frequency $\widetilde{\omega}_{\mathrm{AM}}$, though the actuation induces motion at other frequencies via the intracavity intensity beatnotes (this point is discussed in §4). Thus, the only component of the force vector contributing to the detected motion when $\widetilde{\omega}_{\mathrm{AM}} \approx \widetilde{\omega}_i^{(0)}$ is $\boldsymbol{f}_i$, where $\boldsymbol{f}_1 \propto (1\ 0\ 0)^{\mathrm{T}}, \boldsymbol{f}_2 \propto (0\ 1\ 0)^{\mathrm{T}}, \boldsymbol{f}_3 \propto (0\ 0\ 1)^{\mathrm{T}}$, corresponding to the three uncoupled modes. The motion induced by $\boldsymbol{f}_i$ is the sum of (three) Lorentzians with amplitudes $s_{ij}$.

The amplitude $s_{ij}$ is proportional to the product of the projection of the actuation force $\boldsymbol{f}_i$ onto the $j^{\mathrm{th}}$ left and right eigenvectors of $\mathbf{H}$. In the basis of the uncoupled modes

$$\mathbf{S} = \begin{bmatrix} s_{11} & s_{12} & s_{13} \\ s_{21} & s_{22} & s_{23} \\ s_{31} & s_{32} & s_{33} \end{bmatrix} = \begin{bmatrix} T_{11}(T^{-1})_{11} & T_{12}(T^{-1})_{21} & T_{13}(T^{-1})_{31} \\ T_{21}(T^{-1})_{12} & T_{22}(T^{-1})_{22} & T_{23}(T^{-1})_{32} \\ T_{31}(T^{-1})_{13} & T_{32}(T^{-1})_{23} & T_{33}(T^{-1})_{33} \end{bmatrix}$$

where $\mathbf{T}$ diagonalizes the dynamical matrix, i.e. $\mathbf{H} = \mathbf{TDT}^{-1}$ (see §4). The columns of $\mathbf{T}$ are the right eigenvectors of $\mathbf{H}$ and the rows of $\mathbf{T}^{-1}$ are its left eigenvectors.

We now provide some intuition for why $E = (\det(\mathbf{S}))^{-2}$ vanishes at an EP. In the basis of the system's right eigenvectors, the force $\boldsymbol{f}_i \propto ((T^{-1})_{1i}\ ,\quad (T^{-1})_{2i}\ ,\quad (T^{-1})_{3i})^{\mathrm{T}}$, and in the basis of the system's left eigenvectors, $\boldsymbol{f}_i^{\mathrm{T}} \propto (T_{i1}\ ,\quad T_{i2}\ ,\quad T_{i3})$. At an EP, neither the left nor the right eigenvectors span the full space; therefore, at least two projections of a generic vector (like $\boldsymbol{f}_i$) onto both the left and right eigenvectors must diverge. This implies that at least two columns of $\mathbf{S}$ diverge, and hence that $\det[\mathbf{S}]$ diverges and $E = (\det(\mathbf{S}))^{-2}$ vanishes.

We note that, formally, $E = (\det(\mathbf{S}))^{-2}$ has the undesirable property of depending explicitly on the choice of basis. However, there is a basis that is naturally chosen by our susceptibility measurements (specifically, that of the uncoupled modes), and we will show below that we can still make a generic connection between the vanishing of $E$ and the coalescence of two eigenvectors.

Despite the basis dependence, $E$ also has two nice properties. First, $E$ does not depend on the arbitrary nonzero scale of the eigenvectors because of the multiplication of matrix elements of $\mathbf{T}$ with those of $\mathbf{T}^{-1}$ in the definition of $\mathbf{S}$. Second, taking the square of $\det(\mathbf{S})$ ensures that $E$ does not depend upon the ordering of the eigenvectors, and is single-valued.

We now show that if $\mathbf{T}$ is chosen to vary smoothly away from degeneracies[6] $E$ generically goes to 0 as a second-order EP is approached, while its phase winds by $2\pi$ if such a point is encircled. Without loss of generality, we focus this discussion on a $2 \times 2$ traceless matrix because close to a second-order EP only the two-dimensional subspace spanned by the two coalescing eigenvectors is relevant. For a smooth choice of the eigenvector matrix, its general form is (see Eq. 9 of Ref. [viii])

$$\mathbf{T} \sim \mathbf{V}\begin{pmatrix} 1 & 1 \\ -\sqrt{x} & \sqrt{x} \end{pmatrix} \text{ as } |x| \to 0$$

where $x = \det(\mathbf{H}_0)$ and $x = 0$ at a second-order EP. The first column of $\mathbf{V}$ (i.e., $(V_{11}, V_{21})^T$) defines the unique (right) eigenvector at $x = 0$. All that we know, in general, about the $x$-independent basis transformation matrix $\mathbf{V}$ is that $\det(\mathbf{V}) \neq 0$, which follows from $\det(\mathbf{T}) \neq 0$ for $x \neq 0$. After some algebra, we find:

$$\mathbf{S} \sim \frac{1}{2}\begin{pmatrix} 1 + \frac{V_{11}V_{21}}{\sqrt{x}\det(\mathbf{V})} & 1 - \frac{V_{11}V_{21}}{\sqrt{x}\det(\mathbf{V})} \\ 1 - \frac{V_{11}V_{21}}{\sqrt{x}\det(\mathbf{V})} & 1 + \frac{V_{11}V_{21}}{\sqrt{x}\det(\mathbf{V})} \end{pmatrix} \text{ as } |x| \to 0$$

After some more algebra, this becomes $E \sim x\left(\frac{\det(\mathbf{V})}{V_{11}V_{21}}\right)^2$. This shows that, as long as $V_{11}$ and $V_{21}$ are non-zero, the claimed properties of $E$ follow from $E \sim x$ as $|x| \to 0$. To finish the proof, we argue that $V_{11}$ and $V_{21}$ are generically non-zero. Indeed, for $V_{11}$ or $V_{21}$ to vanish would require fine-tuning all the matrix elements of the traceless matrix $\mathbf{H}_0$ (i.e., six real parameters), while generically we only have access to two real control parameters (which determine $x$).

**§6. Rastering the hypersurface $\mathcal{S}$**

The hypersurface $\mathcal{S}$ described in the main text is the boundary of a 4D hyperrectangle, and is a union of eight 3D hyperrectangles, which we refer to as "faces". Each of these 3D faces is spanned by three components of $\boldsymbol{\Psi}$ (for example, $P_1, P_2, P_3$) which range from their minimum value to their maximum value (given below), while the remaining component of $\boldsymbol{\Psi}$ (in this example it would be $\delta$) is held fixed at either its maximum or its minimum value. As a result, the 3D faces span the ranges:

$-10\text{ kHz} \leq \delta/2\pi \leq 106\text{ kHz}$
$22\ \mu\text{W} \leq P_1 \leq 240\ \mu\text{W}$
$289\ \mu\text{W} \leq P_2 \leq 675\ \mu\text{W}$
$78\ \mu\text{W} \leq P_3 \leq 702\ \mu\text{W}$

For ease of analysis, measurements of $\boldsymbol{\lambda}$ were taken by densely rastering $\boldsymbol{\Psi}$ within sixty-one 2D "sheets", each lying within one of the eight 3D faces. The locations of these sheets are shown in Extended Data Fig. 4, and the actual data sets (from all 61 sheets) are shown in Video 5.

As can be seen from Extended Data Fig. 4, no 2D sheets lie within the two faces having constant $P_1$. This is because the optomechanical model (described in §3) predicts that the $EP_2$ lie only in the other six faces. The absence of $EP_2$ in the two faces with constant $P_1$ was confirmed by measuring $\boldsymbol{\lambda}$ at several hundred locations in these two faces (not shown).

## §7. Data analysis algorithms

Here we describe the algorithms used to locate the EPs in the 2D data sheets. We also describe the processing (outlier rejection and Gaussian filtering) applied to the data in these 2D sheets in order for the algorithms to perform effectively.

As described in the main text, each measurement of a mechanical spectrum (i.e., with the control parameters $\boldsymbol{\Psi}$ set to a specific value) is fit to extract $\boldsymbol{\lambda}$ and $\mathbf{S}$ for this value of $\boldsymbol{\Psi}$. These quantities are then converted into $d, D, E$, and $t$ for this value of $\boldsymbol{\Psi}$. The quantity $t$ is defined as:

$$t \equiv x/y = \left[2\hat{\lambda}_1\hat{\lambda}_2\hat{\lambda}_3/\left(\hat{\lambda}_1^2 + \hat{\lambda}_2^2 + \hat{\lambda}_3^2\right)\right]$$

where $x$ and $y$ are the coefficients of the characteristic polynomial (see Methods and §3.1). The eigenvalues of the traceless version of $\mathbf{H}$ (see §3.1) are $\hat{\lambda}_j = \lambda_j - \bar{\lambda}$, where $\bar{\lambda} = (\lambda_1 + \lambda_2 + \lambda_3)/3$. As described below, $t$ is useful because its complex phase $\theta \equiv \mathrm{Arg}(t)$ provides a coordinate along the trefoil knot.

Much of the analysis used in this study is based on densely rastering two components of $\boldsymbol{\Psi}$, resulting in a 2D "sheet" in which $d(\boldsymbol{\Psi}), D(\boldsymbol{\Psi}), E(\boldsymbol{\Psi})$, and $t(\boldsymbol{\Psi})$ can be displayed. Analyzing the data in these sheets allows for the identification of $EP_2$ and $EP_3$ locations. It is usually straightforward to identify the EP locations from the measured $d(\boldsymbol{\Psi}), D(\boldsymbol{\Psi}), E(\boldsymbol{\Psi})$ as the vanishing or phase-winding of these quantities, which are readily evident in Fig. 2 of the main text and in Video 5.

However, to apply a uniform approach to locating these points, we use a minima-finding algorithm and a vortex-finding algorithm. These algorithms can be adversely impacted by noise in the data and by occasional outlier data points. The noise we refer to is the apparently random pixel-to-pixel variations visible in the top row of Fig. 2 from the main text (i.e., superposed on the smooth behavior that is similar to the corresponding theory plot in the bottom row). The outliers we refer to are the few pixels whose value differs drastically from their neighbors in the top row of Fig. 2 from the main text. As a result, we apply outlier rejection followed by Gaussian filtering to each of the quantities $d(\boldsymbol{\Psi}), D(\boldsymbol{\Psi}), E(\boldsymbol{\Psi})$, and $t(\boldsymbol{\Psi})$, yielding the filtered versions $\bar{d}(\boldsymbol{\Psi}), \bar{D}(\boldsymbol{\Psi}), \bar{E}(\boldsymbol{\Psi})$, and $\bar{t}(\boldsymbol{\Psi})$, which are shown as the middle row in Fig. 2 of the main text and Extended Data Figs. 2,3, as well as Video 5. For the complex quantities $D, E$, and $t$, the real and imaginary parts are treated separately and then recombined.

A minima identification algorithm (described in §7.3) is applied to $\bar{d}$ to locate the $EP_3$ point as described in Methods. Minima identification is also applied to the magnitudes of $\bar{D}$ and $\bar{E}$ to locate the $EP_2$ points in the hypersurface $\mathcal{S}$. A phase-vortex identification algorithm (described in §7.4) is applied to the arguments (i.e. complex phases) of $\bar{D}$ and $\bar{E}$, also to locate the $EP_2$ points in $\mathcal{S}$. Lastly, the argument of $\bar{t}$ at each $EP_2$ is the value of $\theta$ used to color the corresponding point in Figs. 3a,b of the main paper and Extended Data Fig. 5.

The complete data set consisting of the sixty-one 2D sheets used to search for $EP_2$ points in the hypersurface $\mathcal{S}$ is shown in Video 5. The data set of six 2D sheets used to identify the $EP_3$ point is shown in Extended Data Figs 2,3.

**§7.1 *Outlier rejection***

Outliers were identified using a Tukey Fence, which tags a data point at $\boldsymbol{\Psi}$ as an outlier if the value at $\boldsymbol{\Psi}$ is outside the range $\{Q_1 - s \times (Q_3 - Q_1), Q_3 + s \times (Q_3 - Q_1)\}$, where the first and third quartiles $Q_1$ and $Q_3$ are defined over a 5 pixel × 5 pixel neighborhood of $\boldsymbol{\Psi}$ within the 2D sheet under consideration (the neighborhood is clipped for $\boldsymbol{\Psi}$ near the sheet's edge). To ensure that only extreme outliers are tagged, we set $s = 6$. By way of illustration, if the data were Gaussian distributed, this would correspond to tagging only values beyond 8.7 standard deviations.

All tagged outliers are inspected individually to eliminate the possibility of a false tag. The value of each outlier is replaced with the median of its 5 pixel × 5 pixel neighborhood. In the end, ~200 of the ~27,000 measurements (~0.8%) of $\boldsymbol{\lambda}$ and $\mathbf{S}$ on the hypersurface $\mathcal{S}$ were rejected as outliers.

**§7.2 *Gaussian filtering***

After the outlier rejection described above, the data in each 2D sheet is convolved with a 2D Gaussian kernel with HWHM = 1.87 pixels. The filter kernel is clipped (and re-normalized) as appropriate for pixels that lie near the edge of the sheet.

**§7.3 *Minima Identification***

For any quantity $f(\boldsymbol{\Psi})$ (which may be $\bar{d}(\boldsymbol{\Psi})$, $|\bar{D}(\boldsymbol{\Psi})|$, or $|\bar{E}(\boldsymbol{\Psi})|$), a minimum is initially tagged at any value of $\boldsymbol{\Psi}$ at which $f$ is the minimum over its 3 pixel × 3 pixel neighborhood. Since some of these initial tags are caused by noise, we only accept tags at which the magnitude of the second derivative is larger than a specific threshold. In particular, we require $|f''(\boldsymbol{\Psi})| > \zeta$ , where the threshold $\zeta$ is chosen to be $\langle |f''| \rangle + 2\sigma_{|f''|}$, with the mean $\langle \cdots \rangle$ and standard deviation $\sigma$ evaluated over the entire data sheet.

The $\boldsymbol{\Psi}$ that are tagged in this way are reported as the experimentally identified minima. When this analysis is applied to $\bar{d}(\boldsymbol{\Psi})$, the minima correspond to experimental estimates of the $EP_3$ location. When this analysis is applied to $|\bar{D}(\boldsymbol{\Psi})|$ and $|\bar{E}(\boldsymbol{\Psi})|$, the minima correspond to the experimentally identified $EP_2$ points. At each identified minimum of $|\bar{D}|$ and $|\bar{E}|$, the value of $\theta = \mathrm{Arg}(\bar{t})$ at that location is reported as the measured $\theta$ for that $EP_2$. These points are shown in Extended Data Figs 5a,c.

**§7.4 *Phase-Vortex Identification***

For each phase function ($\mathrm{Arg}[\bar{D}]$ and $\mathrm{Arg}[\bar{E}]$), the algorithm starts with a location $\boldsymbol{\Psi}$ within the sheet and then considers the closed counter-clockwise path defined by the eight nearest neighbors of $\boldsymbol{\Psi}$. The point $\boldsymbol{\Psi}$ is tagged as a phase vortex if the unwrapped phase along this closed path changes by $\pm 2\pi$.

It sometimes happens that this approach tags several neighboring $\boldsymbol{\Psi}$ as phase vortices. To determine whether this results from pixelation of the data, or because different portions of the knot actually intersect the sheet in close-by locations, we algorithmically cluster[i] any adjacent points identified as phase vortices based on their value of $\theta$ (which serves to distinguish different parts of the knot from each other). For each cluster identified in this way, the mean value of $\boldsymbol{\Psi}$ is reported as the experimentally identified phase vortex ($EP_2$). Also, the mean value of $\theta(\boldsymbol{\Psi})$ for each cluster is reported as the measured $\theta$ for that $EP_2$. These points are shown in Extended Data Figs 5b,d.

It should be noted that all the phase vortices identified in this work show a winding of $\pm 2\pi$ along the closed path constructed above. This is expected for $\mathrm{Arg}(\bar{D})$ and $\mathrm{Arg}(\bar{E})$, as $\lambda_i(\boldsymbol{\Psi}) \sim |\boldsymbol{\Psi} - \boldsymbol{\Psi}_{\mathrm{EP2}}|^{1/2}$ in the neighborhood of an $EP_2$ point at $\boldsymbol{\Psi}_{\mathrm{EP2}}$.[ix] The proof of this is given in Ref. [i]. See also §5.

### §7.5 *Theory Plots of D and E*

Eqn. S1 (see §3) can be numerically diagonalized at a given $\boldsymbol{\Psi}$ to find the eigenvalues $\boldsymbol{\lambda}(\boldsymbol{\Psi})$. Similarly, we calculate $\mathbf{S}(\boldsymbol{\Psi})$ from the $\mathbf{T}$-matrix associated with this diagonalization, cf. §4. The theoretical $D(\boldsymbol{\Psi})$ and $E(\boldsymbol{\Psi})$ so evaluated are depicted in the bottom rows in Fig. 2 of the main text, in Extended Data Figs 2,3, and in Video 5.

The cyan squares in the theory plots of these figures mark the roots of $D$ (corresponding to $EP_2$), which are found numerically. To make the numerical root-finding tractable, the $EP_2$ degeneracy of $\mathbf{H}$ is cast as the system of equations

$$\mathrm{Re}[D] = 0$$
$$\mathrm{Im}[D] = 0$$

where $D$ is the discriminant of $p_{\mathbf{H}}$.

In producing these theory plots, the parameters used are those obtained as the best-fit parameters for the knot data shown in Figs. 3a,b of the main text:

$$\boldsymbol{g} = 2\pi \times (0.1979, 0.3442, 0.3092)\ \mathrm{Hz}$$
$$\kappa = 2\pi \times 173.84\ \mathrm{kHz}$$

The process by which the knot is fit is described in Methods.

## §8. Visualizing the eigenvalue braids

Here we describe various aspects of representing the eigenvalue braids, both for the experimental data and the theoretical calculations.

### §8.1 *Producing Extended Data Fig. 1*

This subsection describes how Extended Data Fig. 1 was generated. We first describe the calculation of the trefoil knot of $EP_2$ locations shown in Extended Data Fig. 1a. Then we describe the specific form of the control loops in Extended Data Fig. 1a, and the corresponding eigenvalue braids shown in Extended Data Figs. 1b-d.

Extended Data Fig. 1 was generated by considering the two complex coefficients $x$ and $y$ of the characteristic polynomial $p_{\mathbf{H}} = \lambda^3 - y\lambda - x$ for a traceless $3 \times 3$ matrix. As described in the main text (and in Methods), the space spanned by $x$ and $y$ may be viewed as $\mathbb{R}^4$, with the Cartesian coordinates $(\mathrm{Re}(x), \mathrm{Im}(x), \mathrm{Re}(y), \mathrm{Im}(y))^{\mathrm{T}}$ giving a smooth parametrization of all the eigenspectra in the neighborhood of $EP_3$, which is found at $x = y = 0$.

Extended Data Fig. 1a shows a representation of the unit hypersphere $\mathcal{S}^3$ centered at the origin $(0,0,0,0)^{\mathrm{T}}$. Specifically, Extended Data Fig. 1a shows $\mathcal{S}^3$ mapped by a stereographic projection (whose pole is located at $(-1,0,0,0)^{\mathrm{T}}$) to the space $\mathbb{R}^3$ spanned by the Cartesian coordinates $(X, Y, Z)^{\mathrm{T}}$ defined via

$$\mathrm{Re}(x) = \frac{1-X^2-Y^2-Z^2}{1+X^2+Y^2+Z^2} \qquad \mathrm{Im}(x) = \frac{2Z}{1+X^2+Y^2+Z^2} \qquad \mathrm{Re}(y) = \frac{2X}{1+X^2+Y^2+Z^2} \qquad \mathrm{Im}(y) = \frac{2Y}{1+X^2+Y^2+Z^2}$$

The space spanned by $(X, Y, Z)^{\mathrm{T}}$ is the space shown in Extended Data Fig. 1a.

The yellow curve in Extended Data Fig. 1a is defined by two constraints:

$$|x|^2 + |y|^2 = 1$$
$$4y^3 - 27x^2 = 0$$

The first constraint simply defines $\mathcal{S}^3$ and so is satisfied everywhere in the space shown in Extended Data Fig. 1a. The second constraint corresponds to the vanishing of the discriminant of $p_{\mathbf{H}}$ (which is $D = 4y^3 - 27x^2$). Since $D$ vanishes just where two (or more) eigenvalues of $\mathbf{H}$ are degenerate, and since $EP_3$ is not in $\mathcal{S}^3$, the orange curve shows all of the EPs in $\mathcal{S}^3$, and all of these are $EP_2$. This curve is a trefoil knot, as described in Methods.

Each of the three loops (green, red, and blue) shown in Extended Data Fig. 1a can be written in the coordinate system $(X, Y, Z)^T$ as:

$$X(d, r, \theta, \phi, s) = \big((d - r)\sin(\theta) + r\sin(\theta + 2\pi s)\big)\cos(\phi)$$
$$Y(d, r, \theta, \phi, s) = \big((d - r)\sin(\theta) + r\sin(\theta + 2\pi s)\big)\sin(\phi)$$
$$Z(d, r, \theta, \phi, s) = (d - r)\cos(\theta) + r\cos(\theta + 2\pi s)$$

where $\{d, \theta, \phi\}$ denotes the basepoint of the loop in spherical polar coordinates, $r$ is the loop's radius, and $s \in [0,1]$ parameterizes the position along the loop (i.e. $s = 0$ and $s = 1$ both correspond to the loop's basepoint).

All three loops in Extended Data Fig. 1a have the same base point at $(d = 2.5, \theta = \frac{5\pi}{12}, \phi = -\frac{\pi}{12})^T$, which is shown as the black cross. Loops from three distinct homotopy classes were realized by using $r = 0.4, 0.74, 1.2$ for the green, red, and blue loop respectively.

To display the eigenvalue braids produced by each of these loops (shown in Extended Data Figs. 1b-d), we find the three roots of $p_{\mathbf{H}}$ for 101 values of $s$ ranging from 0 to 1. For each value of $s$, the three roots (which comprise the eigenspectrum $\boldsymbol{\lambda}$ of $\mathbf{H}$) are plotted in the complex plane. Their evolution as a function of $s$ is shown in Extended Data Figs. 1b-d by stacking a copy of the complex plane for each value of $s$. The black crosses highlight $\boldsymbol{\lambda}$ at $s = 0$ (the bottom of each plot), which by construction is identical to $\boldsymbol{\lambda}$ at $s = 1$ (the top of each plot).

### §8.2 *Producing Figs. 3c-h of the main paper*

The three control loops shown in Figs. 4c-h of the main text were assembled from data taken in two of the sixty-one 2D sheets. In Figs. 4c-e of the main paper, these loops are shown in the stereographic projection of the hypersurface $\mathcal{S}$ (described in Methods) in order to highlight their relationship with the knot of $EP_2$.

In Extended Data Fig. 10, we show how these three loops lie within their 2D sheets. In Extended Data Fig. 10, each gray disc represents a value of $\boldsymbol{\Psi}$ at which $\boldsymbol{\lambda}$ is measured (i.e., a "pixel" in the 2D sheets of Video 5). The green, red, and blue rectangles show the control loops that are produced by selecting (respectively) 21, 59, and 123 of these pixels. For each of these loops, the discrete variable $\xi$ indexes the pixels along the loop (e.g., $1 \le \xi \le 59$ for the red loop).

To compare the measured braids with theory, Extended Data Fig. 6 shows the same panels as in Fig. 4c-h of the main paper, but together with the $\boldsymbol{\lambda}(\boldsymbol{\Psi})$ calculated using the best-fit parameters from the fit described in Methods (Extended Data Figs. 6g-i).

### §8.3 *Coloring the braid strands*

Figures 3,4 of the main paper (as well as Extended Data Figs. 6,7) show experimentally measured braids. These are realized by stepping the parameters $\boldsymbol{\Psi}$ around the loops shown in Figs. 3c-e and 4a (and Extended Data Figs. 6a-c and 7a-c). At each value of $\boldsymbol{\Psi}$, the spectrum $\boldsymbol{\lambda}$ is determined from measurements as described in §4. The complete set of these measurements (i.e., at all of the values of $\boldsymbol{\Psi}$ around the loop) produces the data points shown in Figures 3f-h and 4b,c of the main paper (and in Extended Data Figs. 6d-f and 7d-f).

However, coloring the individual strands (e.g., as light green, green, and dark green as in Fig. 3f) is potentially ambiguous. This is a consequence of the fact that $\boldsymbol{\Psi}$ is always stepped by a finite amount between measurements, while the components of $\boldsymbol{\lambda}$ at one value of $\boldsymbol{\Psi}$ are associated with specific components of $\boldsymbol{\lambda}$ at some other $\boldsymbol{\Psi}$ only via the fact that $\boldsymbol{\lambda}$ is a smooth function of $\boldsymbol{\Psi}$ (so long as $\boldsymbol{\Psi}$ is not an EP).[ix]

This means that if the steps in $\boldsymbol{\Psi}$ are sufficiently fine (and if the measurements of $\boldsymbol{\lambda}$ have little noise), then the braid strands' identities will be evident from step to step. But if $\boldsymbol{\Psi}$ is stepped too coarsely (or if the measurements of $\boldsymbol{\lambda}$ are very noisy), then it will not be evident how to identify the strands from step to step.

It can be seen from Figs. 3f-h and 4b,c (as well as from the rotating versions of these figures, Video 4) that the steps in $\boldsymbol{\Psi}$ are sufficiently fine (and the noise in $\boldsymbol{\lambda}$ sufficiently small) that it would be straightforward to connect the measurements of $\boldsymbol{\lambda}$ into three braids "by eye". However, to avoid any potential ambiguity, we implemented this "coloring" of the strands using a simple algorithm. Specifically, with each increment of $\xi$ (which indexes the position along the control loop) i.e., from $\xi$ to $\xi + 1$, each component of $\boldsymbol{\lambda}(\xi + 1)$ is associated with a component of $\boldsymbol{\lambda}(\xi)$ such that the sum of the distances

$$Q = \sum_{\boldsymbol{m}=\mathbf{1}}^{\mathbf{3}} |\lambda_{\boldsymbol{m}}(\xi + 1) - \lambda_{\boldsymbol{m}}(\xi)|^2$$

is minimized. More precisely, $Q$ is minimized over the six possible choices for identifying the components of $\boldsymbol{\lambda}(\xi + 1)$ with those of $\boldsymbol{\lambda}(\xi)$. Repeating this process for every value of $\xi$, the braids are tracked and colored as depicted in Figs. 3,4 of the main text and in Extended Data Figs. 6,7.

## §9. Videos

**Video 1: Laying out the hypersurface $\mathcal{S}$ in terms of the experimental parameters.** The surface of a 4D hyperrectangle is a union of eight 3D hyperrectangles. These 3D hyperrectangles are connected to each other via their common 2D faces. As described in Methods, the "rectilinear stereographic" projection (used in Fig. 3b of the main paper) "glues" those common 2D faces together in a way that is isomorphic to the standard stereographic projection. This video illustrates the construction of the rectilinear stereographic projection from the eight 3D data sets.

**00:00 – 00:07** An arrangement of the eight 3D hyperrectangles is shown. Each hyperrectangle is labelled by its fixed control parameter (green text). The labelled arrows on the axes of each hyperrectangle indicate the remaining three control parameters which vary within their bounds, i.e.: $-10$ kHz $\leq \delta/2\pi \leq 106$ kHz, $22$ μW $\leq P_1 \leq 240$ μW, $289$ μW $\leq P_2 \leq 675$ μW, $78$ μW $\leq P_3 \leq 702$ μW.

**00:07 – 00:15** Some pairs of common 2D faces are chosen and highlighted in different colors.

**00:16 – 00:18** The highlighted faces are glued together by translating them towards the $P_1 = 22$ μW hyperrectangle.

**00:18 – 00:24** The remaining hyperrectangles are glued on their common 2D faces. Then the six nearest neighbors of the $P_1 = 22$ μW hyperrectangle are subject to a bilinear transformation that stretches them transverse to their $P_1$ axis. The eighth and final hyperrectangle ($P_1 = 240$ μW) is turned inside-out and its contents are stretched over the region exterior to the rest of the face (i.e., extending to infinity). The wireframe figure at 00:24 is the end result of the rectilinear stereographic projection.

**Video 2: Visualizing the $EP_2$ knot in the rectilinear stereographic projection.** This video shows how the measured $EP_2$ locations (and the best-fit knot) appear in each of the eight 3D faces of the hypersurface $\mathcal{S}$. It also shows the smooth transformation of these faces (along with the data & fit) into the rectilinear stereographic projection of Fig. 3b from the main text.

**00:00 – 00:05** The eight 3D faces of $\mathcal{S}$ are shown. They are arranged to form a "net" of the hypersurface $\mathcal{S}$. Also shown are the measured $EP_2$ locations (colored circles) and the best-fit knot (solid curve) from Figs. 3a,b of the main text. For each 3D face, the green text indicates the control parameter that is held constant.

**00:05 – 00:16** The net is rotated to give a complete view. Note that the two 3D faces in which $P_1$ is held constant do not contain any $EP_2$ locations (in either the data or the fit).

**00:16 – 00:26** The eight 3D faces, along with the data and fit inside them, are continuously deformed to realize the rectilinear stereographic projection. Note that the two empty 3D faces (i.e., the ones with constant $P_1$) are mapped to the innermost cube and to the region outside the frame. Thus, all of the $EP_2$ locations (in both the data and the fit) lie in the six hexahedrons that surround the central cube.

**00:26 – 00:37** The rectilinear stereographic projection is rotated to give a complete view.
**00:37 – 00:40** The axis labels are added.

**Video 3:** This video is simply a rotating version of Figs. 3a,b from the main text.

**Video 4:** This video is simply a rotating version of Figs. 3c-h from the main text.

**Video 5: The sixty-one 2D data sheets used to locate the $\boldsymbol{\Psi}_{\mathrm{EP2}}$ in the hypersurface $\boldsymbol{\mathcal{S}}$.** Each panel shows the complex-valued quantities *D* and *E* measured on a 2D sheet in $\mathcal{S}$. From left to right, the columns show $\mathrm{Abs}(D)$, $\mathrm{Arg}(D)$, $\mathrm{Abs}(E)$, and $\mathrm{Arg}(E)$. In each sheet, two of the control parameters are held fixed (these fixed values are given in the upper right corner). The other two control parameters are scanned, and form the horizontal and vertical axes of the 12 panels. The top row shows the raw data. The middle row shows the data after outlier rejection and convolution with a Gaussian (described in §7). The cyan circles are the $\boldsymbol{\Psi}_{\mathrm{EP2}}$, which are identified algorithmically (also described in §7). The bottom row shows *D* and *E* as calculated from optomechanics theory. The cyan squares are the $\boldsymbol{\Psi}_{\mathrm{EP2}}$ determined from this calculation. In addition, each panel includes two views of the $\boldsymbol{\Psi}_{\mathrm{EP2}}$ data and best-fit knot (as in Fig. 3a,b of the main text) in which the specific $\boldsymbol{\Psi}_{\mathrm{EP2}}$ found in that panel are shown in red.

## §10. Table

| Parameter 1 → / Sheet ↓ | $\delta/2\pi$ (kHz) ↓ | $P_1$ (μW) ↓ | $P_2$ (μW) ↓ | $P_3$ (μW) ↓ |
|---|---|---|---|---|
| Parameter 1 × $\delta$ | -- | 134 | 431 | 293 |
| Parameter 1 × $P_1$ | 47.8 | -- | 429 | 321 |
| Parameter 1 × $P_2$ | 52.5 | 122 | -- | 299 |
| Parameter 1 × $P_3$ | 61.9 | 124 | 425 | -- |
| | | | | |
| Mean | 54.1 | 128 | 428 | 304 |
| Std. Dev. | 7.2 | 8 | 3 | 15 |

**Table 1**
Location of the minimum in $d$ for each of the six 2D sheets shown in Extended Data Figs. 2,3. See Methods.

## §11 Supplementary References